\documentclass{article}
\pdfoutput=1
\usepackage{microtype}
\usepackage{graphicx}
\usepackage{booktabs} % for professional tables
\usepackage{caption}
\usepackage{subcaption}
\usepackage{hyperref}

\usepackage[accepted]{icml2024}

\usepackage{amsmath}
\usepackage{amssymb}
\usepackage{mathtools}
\usepackage{amsthm}

\usepackage[capitalize,noabbrev]{cleveref}

\theoremstyle{plain}

\theoremstyle{definition}

\theoremstyle{remark}

\usepackage[textsize=tiny]{todonotes}

\usepackage{xspace}
\newcommand{\name}{NeWRF\xspace}

\icmltitlerunning{NeWRF: A Deep Learning Framework for
Wireless Radiation Field Reconstruction and Channel Prediction }

\begin{document}

\twocolumn[
\icmltitle{NeWRF: A Deep Learning Framework for \\
Wireless Radiation Field Reconstruction and Channel Prediction}




\begin{icmlauthorlist}
\icmlauthor{Haofan Lu}{ucla}
\icmlauthor{Christopher Vattheuer}{ucla}
\icmlauthor{Baharan Mirzasoleiman}{ucla}
\icmlauthor{Omid Abari}{ucla}
\end{icmlauthorlist}

\icmlaffiliation{ucla}{Department of Computer Science, University of California Los Angeles (UCLA), Los Angeles, United States}

\icmlcorrespondingauthor{Haofan Lu}{haofan@cs.ucla.edu}
\icmlkeywords{Neural Radiance Fields, Wireless Communication, Wireless Network Optimization}

\vskip 0.3in
]



\printAffiliationsAndNotice{}  % leave blank if no need to mention equal contribution

\begin{abstract}
We present \name, a novel deep-learning-based framework for predicting wireless channels. Wireless channel prediction is a long-standing problem in the wireless community and is a key technology for improving the coverage of wireless network deployments. Today, a wireless deployment is evaluated by a site survey which is a cumbersome process requiring an experienced engineer to perform extensive channel measurements. To reduce the cost of site surveys, we develop \name, which is based on recent advances in Neural Radiance Fields (NeRF). \name trains a neural network model with a sparse set of channel measurements, and predicts the wireless channel accurately at any location in the site. We introduce a series of techniques that integrate wireless propagation properties into the NeRF framework to account for the fundamental differences between the behavior of light and wireless signals. We conduct extensive evaluations of our framework and show that our approach can accurately predict channels at unvisited locations with significantly lower measurement density than prior state-of-the-art. 
\end{abstract}

\section{Introduction}
\label{sec:intro}
Wireless networks (such as WiFi and 5G) have become an essential part of our lives. Real-world WiFi and cellular deployments commonly encounter many issues such as dead spots, dropped signals, sudden outages, and slow throughput, potentially undermining the connectivity of wireless devices. To solve these issues, site surveys~\cite{site_survey} are often conducted to measure the quality of the wireless network in various locations and, if required, optimize the deployment of base stations. However, to conduct an effective site survey, an experienced engineer needs to perform measurements for a very dense grid of points. Today, this exhaustive measurement is rarely done due to its intractable time and cost. Instead, a sparse measurement approach is usually taken, where the engineer walks with a radio receiver in the site and measures the wireless channels at random locations. A wireless channel is a measure of the distortions imposed on wireless signals as they propagate from a transmitter to a receiver, which determines the quality of communication. The wireless channel is represented as a complex number that characterizes factors such as signal attenuation, phase rotation, and interference. Although the random sparse measurement approach is much more efficient than the exhaustive grid-based survey, it fails to uncover the signal quality at unvisited locations; therefore, can potentially miss many dead spots. As a supplement, wireless ray-tracing simulations~\cite{remcom_wirelessinsite} are also performed to analyze the distribution of wireless fields, which uses Computer-Aided Design (CAD) models of the environment. However, ray-tracing simulations are not reliable since CAD models cannot fully replicate how real-world environments interact with wireless signals. Therefore, successive site surveys are always required for further validation and calibration.
 \begin{figure}[t]
    \centering
    \includegraphics[width=0.8\columnwidth]{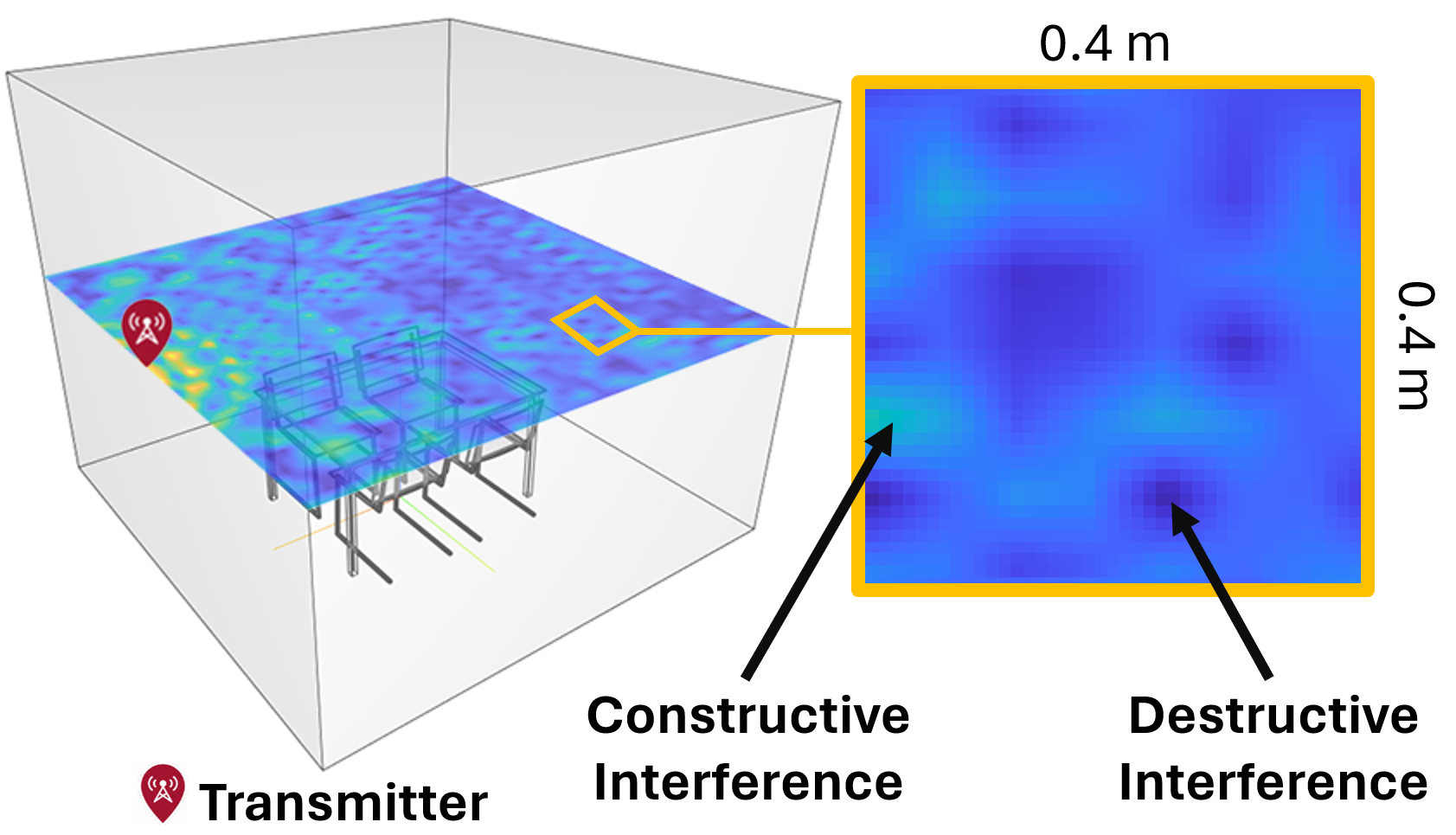}
    \caption{\textbf{The tomography of a wireless field (channel) in a simple environment.} Like water ripples, wireless signals can add up constructively or destructively, causing fine-scale spatial variation. Our algorithm, \name, reconstructs this field from a sparse set of channel measurements.}
    \label{fig:theme}
    \vspace{-10pt}
\end{figure}

In this work, we introduce a deep learning framework that reduces the time and cost of wireless site surveys. Our goal is to predict the wireless channel in every location in the space only using a sparse set of channel measurements. One straightforward approach to accomplish this is to train a machine-learning model that maps a spatial location to the corresponding wireless channel. However, this approach fails to incorporate the underlying physics of wireless propagation, and thus still requires very dense measurements to achieve reasonable performance. The challenge stems from the fine-scale spatial variation of the channel, caused by the constructive/destructive interference of electromagnetic waves propagating via different paths (as shown in Figure~\ref{fig:theme}).

To solve this problem, we are inspired by recent advances in computer vision and computer graphics, in particular, Neural Radiance Fields (NeRF)~\cite{mildenhall2021nerf}. NeRF is a novel deep learning framework that achieves impressive performance in 3D reconstruction and view synthesis tasks. Specifically, by learning from a collection of photos of a scene, NeRF can synthesize photorealistic images from any viewpoint, even those not captured in the original dataset. The key innovation of NeRF is to represent a scene's radiance field as a continuous function, parameterized by a Multilayer Perceptron (MLP) network. This model learns to correlate 3D spatial coordinates with colored radiance and density from 2D images. NeRF uses ray-tracing and volume-rendering techniques to synthesize images, which incorporates the underlying physics of light propagation. 

In this work, we build on NeRF to predict wireless channels at unvisited locations by learning the wireless radiation scenes from a sparse set of channel measurements. However, adapting NeRF to predict wireless channels requires addressing several important challenges due to fundamental differences between visible light and wireless signals. First of all, wireless environments, typically at the scale of a room, are much larger and more complicated than the scenes handled in NeRF. Although there have been recent attempts to learn room-scale scenes with NeRF~\cite{roessle2022dense,azinovic2022neural}, they all require additional depth information for regularization, which is hard to obtain for wireless measurements. Second, each wireless measurement results in a single complex number, which contains much less information than an image of thousands of pixels. As a consequence, training a model with merely wireless channel measurements suffers from restricted information. Third, unlike image sensors in a camera that capture light rays from a single direction, wireless antennas capture signals from all directions. Tracing rays for antennas requires sampling all directions, which significantly enlarges the search space. We find that training the model to trace all ray directions is both memory-hungry and results in poor convergence. Last but not least, the propagation of wireless signals is much more complex than that of visible light. Effects such as attenuation, phase rotation, reflection, and interference need to be characterized carefully to enable the model to learn a realistic wireless radiation scene.

To address these challenges, we make the following contributions:

\begin{itemize}
\itemsep0em 
    \item We propose \name, the first NeRF-based wireless channel prediction framework that integrates wireless propagation characteristics into NeRF, enabling the learning of wireless radiation scenes from sparse sets of channel measurements.
    \item We improve the convergence of our model through a novel ray-casting scheme that uses direction-of-arrival measurements from a wireless antenna array to guide the model to search the most critical directions. 
    \item By inspecting the learnt model of \name, we discover, for the first time, the simple nature of wireless scenes that enables our model to learn, even in complex room-scale spaces using only sparse channel measurements.
    \item We propose a novel ray-searching algorithm to estimate the direction-of-arrival, enabling accurate channel predictions at any unvisited location.
\end{itemize}

In summary, we propose \name, a novel deep learning framework for predicting wireless channels. 
We believe our work represents the first step towards using NeRF to predict wireless channels in complex indoor environments with sparse channel measurements, opening up many new opportunities to optimize the performance of future wireless networks. The datasets and code are available at: 
\href{https://github.com/LuHaofan/NeWRF}{https://github.com/LuHaofan/NeWRF}

\section{Preliminaries}
\label{sec:bg}
In this section, we provide background knowledge on both wireless communication and Neural Radiance Fields. 
\begin{figure}[ht]
    \centering
    \includegraphics[width=0.8\linewidth]{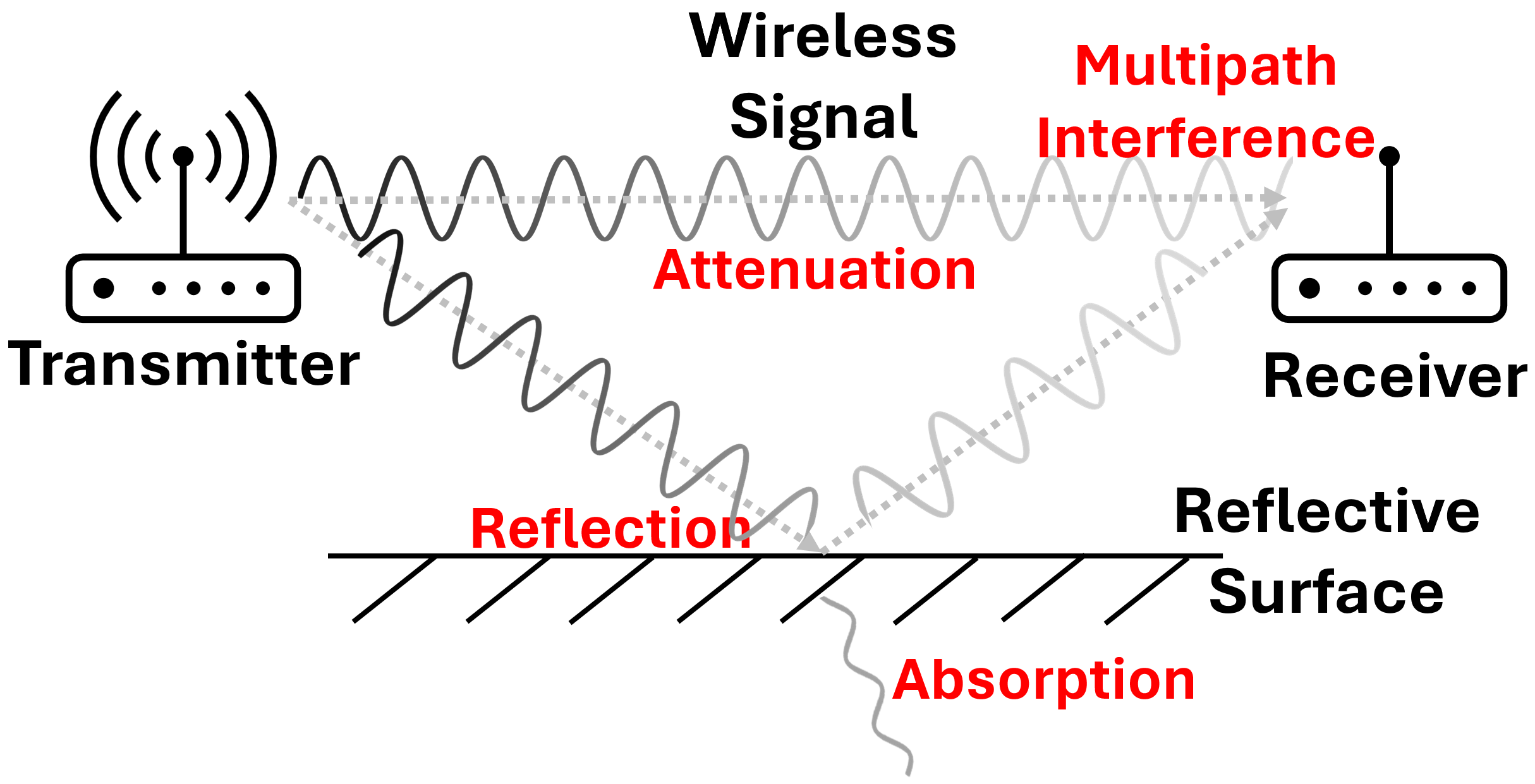}
    \vspace{-5pt}
    \caption{\textbf{Wireless communication model:} Wireless signals emitted by the transmitter experience attenuation, reflection, absorption, multipath interference, etc. before reaching the receiver. Our framework models all of these factors.}
    \label{fig:wireless_comm}
\end{figure}

\noindent\textbf{Wireless Channel}
A wireless communication system consists of a transmitter and a receiver, illustrated in Figure~\ref{fig:wireless_comm}. The transmitter emits wireless signals, the amplitude and phase of which are carefully adjusted to encode data. The transmitted signal can be represented as a complex number $x = Ae^{j\psi}$, where $A$ is the amplitude, and $\psi$ is the phase. As the signal propagates, its amplitude is attenuated by an attenuation factor, $A_{att}$, and the phase is rotated by $\Delta \psi$. The signal captured by the receiver $y$ can be written as 
\begin{equation}
    y=Ae^{j\psi}\times A_{att}e^{j\Delta \psi} = A\cdot A_{att}e^{j(\psi+\Delta \psi)}.
\end{equation}

For free space propagation, i.e. with no objects blocking the propagation path, the amplitude attenuation is proportional to the inverse of the propagation distance, and phase rotation is proportional to the propagation distance~\cite{molisch2012wireless}: 
\begin{equation}
\label{eq:amp_and_phs}
    A_{att}(d) = \frac{c}{4\pi d f},\; \Delta \psi(d) = -2 \pi f d/c.
\end{equation}
where $d$ is the propagation distance, $f$ is the signal frequency, and $c$ is the speed of light. Interactions with objects, for example, reflection and penetration introduce additional attenuation and phase rotations to the signal.

Additionally, transmitted signals propagate via multiple paths to reach the receiver. As a result, the received signal is the sum of multiple copies of the transmitted signal each with different attenuation factors and phase rotations: 
\vspace{-5pt}
\begin{equation}
    y=Ae^{j\psi}\times \sum_{l=0}^{L-1} A_{att}^l e^{j\Delta \psi^l} 
\end{equation}
where $L$ is the total number of propagation paths, and $l$ is the path index. The wireless channel, $h$, is defined as the ratio of the received signal and transmitted signal, and characterizes the distortions introduced by the environment: 
\vspace{-5pt}
\begin{equation}
    h = \frac{y}{x} =\sum_{l=0}^{L-1}A_{att}^le^{j\Delta \psi^l}.
    \label{eq:channel}
\end{equation}
\vspace{-5pt}

\noindent\textbf{Neural Radiance Fields (NeRF).}
NeRF learns a 3D volumetric representation of the radiance field of a scene from 2D images. It represents a scene using an MLP model with 5-dimensional inputs: $f: (x,y,z,\theta,\phi) \longrightarrow (r,g,b,\sigma)$, where $x$, $y$, $z$ are the 3D spatial coordinates, and $\theta, \phi$ are the 2D view direction of a ray traced from the camera. This function translates the 5D coordinates to the radiance of red ($r$), green ($g$), and blue ($b$) light, as well as the volume density, $\sigma$, which describes the degree in which the environment blocks or attenuates light at a location. To generate images from a particular viewpoint, NeRF performs ray-tracing from each pixel of the image, takes discrete sample points $\{\mathbf{p_0}, \mathbf{p_1}, ..., \mathbf{p_N}\}$ along each ray $\mathbf{r}$, and query the MLP model with the coordinates of these samples for the radiance and volume density predictions. The volume density is translated to transmittance, $T$, and opacity, $\alpha$, of the corresponding volume block. Transmittance refers to the amount of light that passes through volume blocks without being absorbed or scattered. While opacity describes the extent to which the light is absorbed or scattered by a particular volume block.
\begin{equation}
    T_i = \exp(-\sum_{j=1}^{i-1}\sigma_j \delta_j),\quad \alpha_i = 1-\exp(-\sigma_i \delta_i),
    \label{eq:transparency}
\end{equation}
where $i$ is the index of sample point, $\delta_i = ||\mathbf{p_{i+1}}-\mathbf{p_i}||$ is the distance between consecutive samples. 

A volume rendering algorithm is then employed to approximate the pixel value, $\hat{C}(\mathbf{r})$, as the weighted sum of the radiance at the sample points:

\begin{equation}
    \hat{C}(\mathbf{r})=\sum_{i=1}^N T_i \alpha_i \mathbf{c}_i,
\end{equation}

where $N$ is the number of sample points along each ray, and $\mathbf{c}_i=[r,g,b]$ is the vector of radiance emitted at the sample point $\mathbf{p_i}$. The generated image is compared with the ground truth image to calculate the loss and update model parameters during the training.

\section{Related Work}
\label{sec:related}
NeRF has recently received huge attention from the computer vision and graphics communities. There has been a lot of work trying to improve and extend the original NeRF in various regards. For instance, ~\citet{martin2021nerf} extends NeRF to synthesize large scenes from unconstrained online photo collections; ~\citet{pumarola2021d} enables NeRF to learn dynamic scenes; ~\citet{mildenhall2022nerf} handles raw images taken in dark environments. These works all use NeRF in the visible light spectrum which is fundamentally different from wireless signals. In contrast, we address several challenges and extend the NeRF formulation to the radio frequency spectrum to learn the radiation scenes of wireless signals. 

Predicting the wireless channel has been a long-standing problem in the wireless community~\cite{karanam2022foundation, malmirchegini2012spatial,krijestorac2021spatial}.
The existing studies are mainly divided into two categories: temporal prediction and spatial prediction. Temporal prediction work focuses on predicting the wireless channel at a future timestamp given the past observations~\cite{Formis_2023,Rajat2023}. Spatial prediction work attempts to predict the channel at an unseen location given the measurements at some other locations~\cite{karanam2022foundation}. In this work, we focus on solving the spatial prediction problem.

There have been some pioneer attempts in this field; however, the fine-scale spatial variation of channels makes this task extremely challenging. 
\citet{karanam2022foundation} proposes an analytical method to predict the wireless channel by extrapolating measurements taken at the boundaries of the area of interest. However, their method cannot deal with complex indoor environments, where multipath interference is more severe. Also, it is limited to 2D receiver layouts with carefully designed boundaries. In contrast, our method can predict wireless channels in 3D indoor environments with sparse measurements at random locations. 
A recent work, NeRF\textsuperscript{2}~\cite{nerf2}, proposed a NeRF-inspired technique to predict the direction-of-arrivals heatmap for a fixed wireless receiver when the transmitter is placed at different locations in the space. We note that the varying location of the transmitter results in a varying wireless field at each measurement, which deviates from the nature of NeRF, that captures a static field representation\footnote{This is analogous to learning a NeRF where the lighting condition of the scene changes from one image to another.}. As a consequence, NeRF\textsuperscript{2} needs to take the transmitter locations as a direct input to the MLP model and essentially learns a direct mapping from transmitter locations to the receiving heatmaps. This is evident by their requirement of extremely dense measurements (178.1 $measurements/ft^3$ on average in their datasets) for training. We believe, the reverse problem, i.e. learning a static radiation field sourcing from a fixed transmitter, is more challenging. In this work, we bridge this gap and devise a novel algorithm to learn a wireless radiation field with significantly lower measurement density (i.e. 0.2 $measurements/ft^3$).

Another work~\cite{orekondy2022winert} proposes a NeRF-based neural surrogate for wireless ray-tracing simulation. 
Their method requires knowledge of the environment geometry (given as a CAD model) and focuses on learning the ray-surface interaction behavior for a wireless signal propagating in space. In contrast, our method directly learns the radiation field of the scene and requires no knowledge about the environment geometry (except for the boundary of the scene). This enables us to eliminate the need for CAD models as a replica of the real world and thus can be easily extended to predict wireless channels in real-world scenarios.

\section{\name}
\label{sec:method}
% Overview
In this section, we present the design of \name. We first provide a concrete formulation of the problem in Section~\ref{sec:prob_def}. Then we present our customized neural channel synthesis algorithm in Section~\ref{sec:prop_model}. We improve the convergence of our model with a novel DoA-guided ray-casting method introduced in Section~\ref{sec:ray_casting}. We present the optimization process of \name in Section~\ref{sec:training-pipeline}. In Section~\ref{sec:wireless_scenes}, we uncover the nature of wireless scenes and provide insights on how it enables our model to reconstruct wireless fields in complex environments. Finally, in Section~\ref{sec:ray_searching}, we present our novel ray-searching algorithm for predicting wireless channels at inference time.

\subsection{Problem Description}
\label{sec:prob_def}
\begin{figure}[t]
    \centering
    \includegraphics[width=0.75\linewidth]{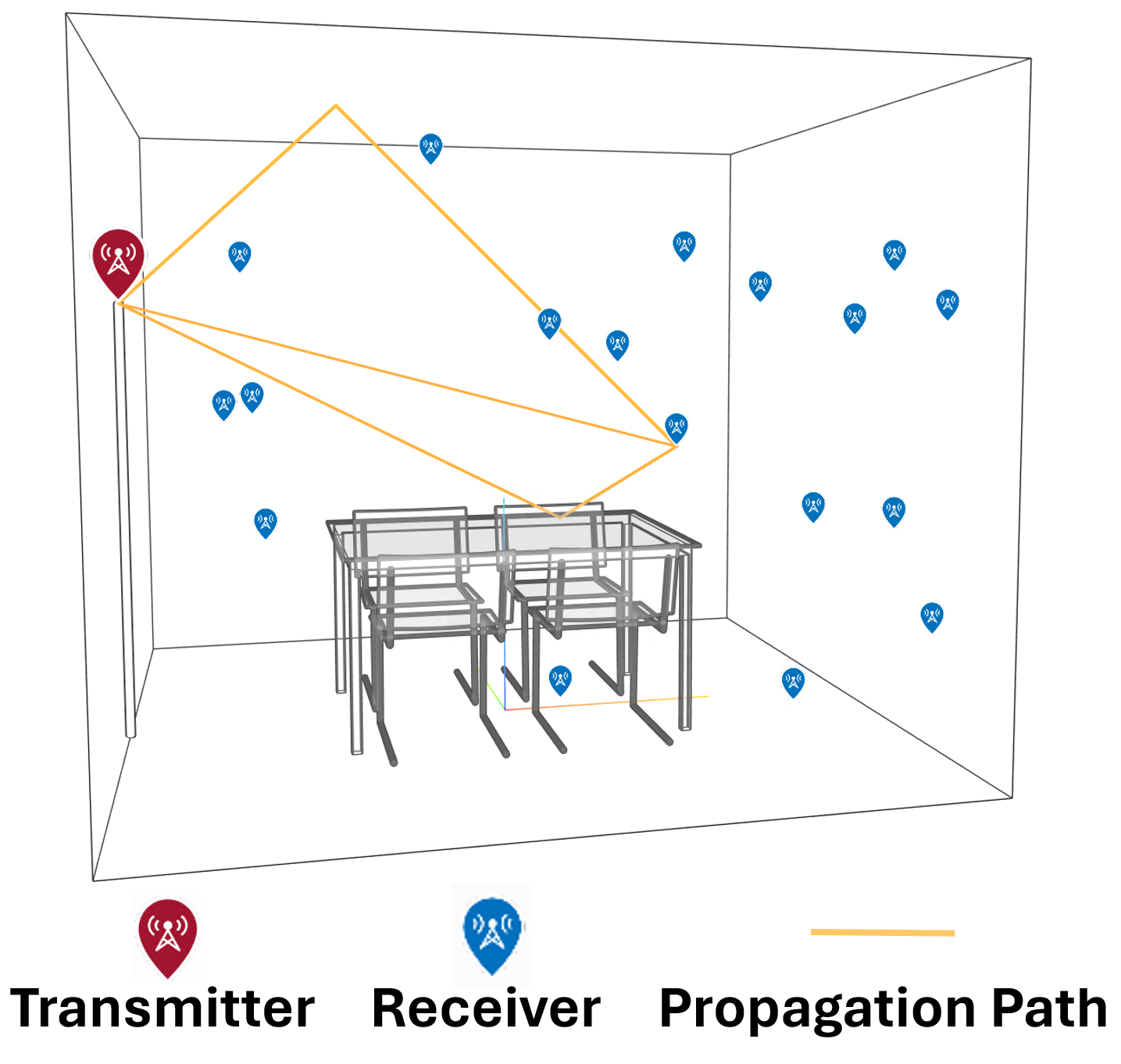}
    \caption{\textbf{An example environment~\cite{matlab_conference} and locations of the transmitter and receivers.} We fix the location of the transmitter and measure the wireless channel at various receiver locations. The wireless channel is the sum of multiple propagation paths from the transmitter to the receiver.}
    \label{fig:conference_room}
    \vspace{-10pt}
\end{figure}

We consider indoor environments (such as the conference room shown in Figure~\ref{fig:conference_room}) with a single transmitter at a fixed location and multiple receivers placed randomly throughout the space. This setting represents a realistic scenario where a WiFi access point talks to multiple client devices, such as mobile phones, laptops, and various Internet-of-Things (IoT) devices. Each receiver (client device) measures the wireless channel and reports its measurements to a central database. Note that when there are multiple transmitters co-exist in the same space, communication systems naturally perform time/frequency multiplexing to avoid interference. Therefore, we assume the receivers can connect to each transmitter in a round-robin manner and construct a wireless scene for each transmitter in such cases. The propagation of wireless signals can be modeled as rays that are emitted from the transmitter, reflected by objects and walls and finally captured by the receiver~\cite{mittra2016computational}. The reflection of wireless signals follows the law of reflection, i.e. the angle of the reflected ray is equal to the angle of the incident ray, since the wavelengths of wireless signals (e.g. 12~cm for 2.4~GHz WiFi signals) are typically much larger than the roughness of object surfaces~\cite{mittra2016computational}. 
Our goal is to predict the wireless channel at any location in the environment using the sparse measurements obtained from the receivers.

\subsection{Neural Wireless Channel Synthesis}
\label{sec:prop_model}
\begin{figure}[t]
    \centering
    \includegraphics[width=0.8\linewidth]{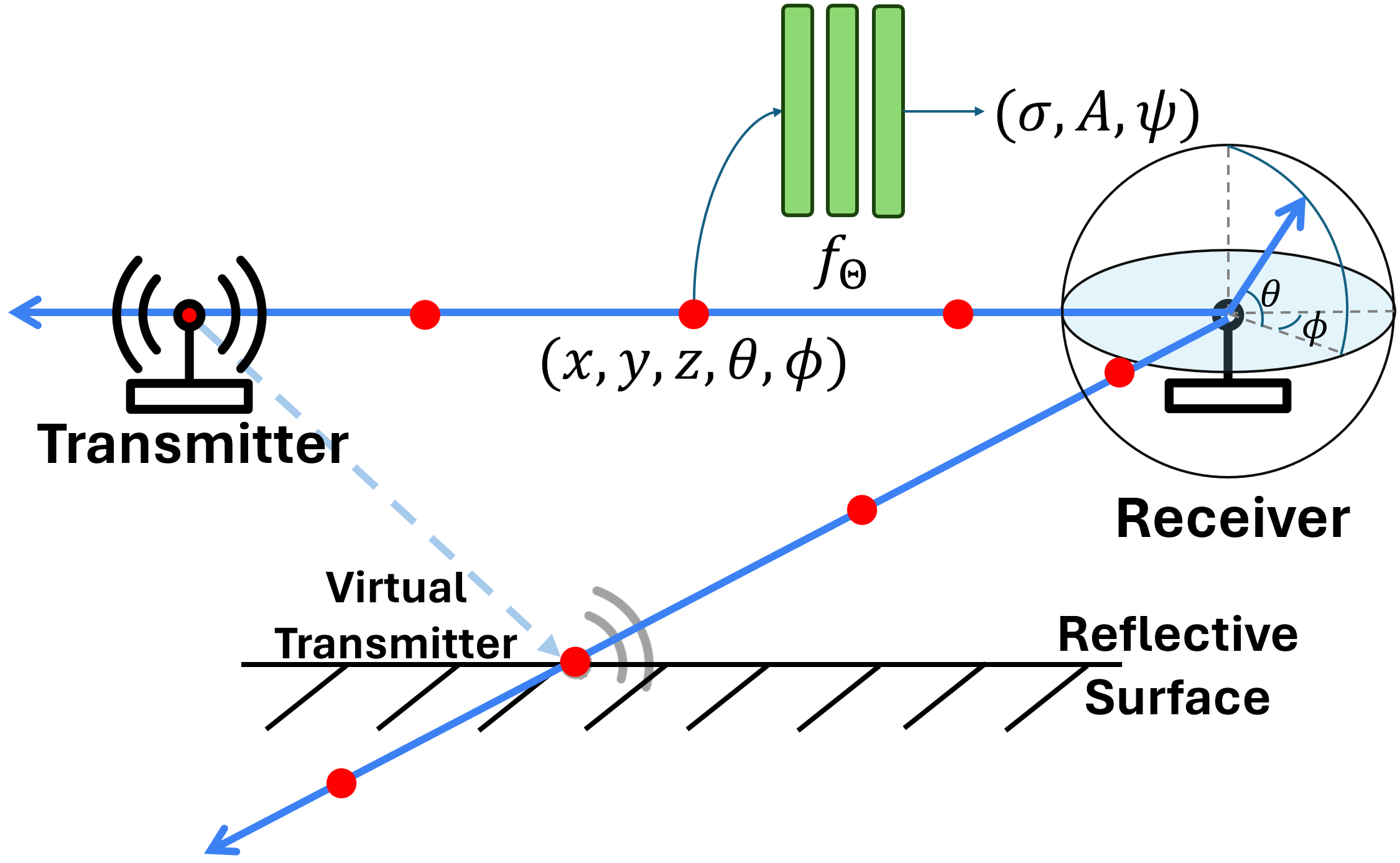}
    \caption{\textbf{An overview of our neural channel synthesis algorithm:} We backtrace each propagation path from a receiver and query an MLP with the 5D coordinates (location and viewing direction) of points along the path, to predict the radiated signal (amplitude and phase) at that point. Then we integrate signals from all paths to synthesize a channel. We model reflections as virtual transmitters at the point of reflection.}
    \label{fig:prop_model}
    \vspace{-18pt}
\end{figure}
The first step to synthesizing the wireless channel is learning a wireless radiation scene. A wireless radiation scene is a continuous function that maps a point in space to the wireless signal emitted from the point. This function is also direction-dependent, since a point may emit signals of different amplitudes and phases towards different directions. We approximate this wireless radiation scene with an MLP model which takes a 5D input of spatial $(x, y, z)$ and direction $(\theta, \phi)$ coordinates. The model outputs $(A, \psi, \sigma)$, which are radiated signal amplitude, signal phase, and volume density, respectively. Volume density is an indicator of the probability of a point being a wireless radiator. 

Figure~\ref{fig:prop_model} illustrates the synthesis process of a wireless channel for a particular receiver location. We backtrace the rays captured by the receiver, and along each ray $\mathbf{r}$, we take multiple samples to obtain a discrete set of points~\footnote{To avoid confusion, in this paper, the term "samples" is used exclusively for the points taken along the rays. We refer to the channel measurements data samples in our datasets with the term "measurements".}. We query the MLP with the spatial and direction coordinates for each sample point, $\mathbf{p}$ to get the radiated signal $A_{\mathbf{rp}}e^{j\psi_{\mathbf{rp}}}$ and the corresponding volume density $\sigma$. We translate the volume density into a weight $w_\mathbf{rp} = T_\mathbf{rp}\alpha_\mathbf{rp}$, which characterizes the contribution of the signal radiated at $\mathbf{p}$ to the channel at the receiver. We follow the definition of $T$ and $\alpha$ in Eq.~\eqref{eq:transparency}. 

Note that two kinds of points contribute most to the channel at the receiver. One is the transmitter itself, which directly acts as a source of radiation, and the other one is any point that reflects the transmitter signal. We consider the latter as virtual transmitters. 
The radiated signal from a virtual transmitter includes the effects of absorption, attenuation, and phase rotation due to the reflection. The effects of free space propagation (Eq.~\eqref{eq:amp_and_phs}) are then explicitly included in our synthesis algorithm. 
Hence the channel at a receiver can be presented as the following equation, where it is calculated by summing up signals radiated from all sample points along all rays: 
\begin{equation}
    \label{eq:rendering}
    \hspace{-2mm}H(f)=\sum_{\mathbf{r}\in \mathbb{L}} \sum_{\mathbf{p}\in \mathbb{P}} w_{\mathbf{rp}} A_{\mathbf{rp}} \frac{c}{4\pi d_\mathbf{p} f} e^{j (\psi_{\mathbf{rp}}- 2\pi f d_\mathbf{p}/c)},
\end{equation}
where $\mathbb{L}$ is the set of rays, and $\mathbb{P}$ is the set of points taken along a ray.

\subsection{Improving the Convergence of \name}
\label{sec:ray_casting}
One key difference between wireless measurements and images is that wireless antennas capture signals from all directions. Thus, in wireless, finding the ray direction to trace is a non-trivial problem. One potential solution is to search exhaustively along a discrete set of azimuth and elevation angle combinations that span all directions. However, this approach has two fatal drawbacks. First, the actual ray directions lie in a continuous space. Approximating them with a discrete set of ray directions would require an extremely dense grid of angles, which significantly increases the memory consumption for training the model. Secondly, for each receiver location, there are only 10$\sim$20 ray directions contributing significantly to the channel measurement, whereas a search grid of 1$^\circ$ resolution involves $360\times180 = 64,800$ ray angles. As a result, optimizing the model to find the actual ray directions from such a large ray space is very difficult, if not impossible. We find through experiments that the model fails to converge with such a large search space of ray directions. To solve this issue, we leverage the fact that the Direction-of-Arrivals (DoA) of wireless signals can be measured using an antenna array at the receiver location~\cite{foutz2022narrowband}. We use DoA measurements as prior knowledge and let the learning algorithm trace rays primarily toward directions identified by the DoA measurements. 
Through our experiments, we find that this DoA-guided ray-casting approach significantly reduces the training time and memory consumption, as well as enables the learning algorithm to converge quickly.

\subsection{Optimizing \name}
\label{sec:training-pipeline}
As in NeRF, We simultaneously optimize two MLP models for each environment. One "coarse" and one "fine". We use mini-batch gradient descent to train our models where at each iteration, we randomly pick a batch of receiver locations from the training set and obtain the ground truth channel and DoA measurements. We cast rays toward each DoA and use the stratified sampling strategy~\cite{mildenhall2021nerf} to take coarse-grained samples uniformly distributed on each ray. We use these samples to query the "coarse" model and use the procedures described in Section~\ref{sec:prop_model} to synthesize coarse predictions of the channel for each receiver location. We use the predicted weights $w_\mathbf{p}$ of coarse samples to perform a round of hierarchical sampling which densely samples regions with high weight values. We query the "fine" model with the union of coarse and fine samples and synthesize the fine channel predictions. The coarse and fine channel predictions are each compared with the ground truth to calculate loss and update the parameters of each model. 

In practice, we let the model predict the real and imaginary parts of the radiated signal, $(I, Q)$, instead of amplitude and phase, $(A, \psi)$. This is because phase is modulo against $2\pi$, which is not differentiable. $A$ and $\psi$ can be easily translated from $I$ and $Q$ using Euler's formula: 
\begin{equation*}
    A = \sqrt{I^2+Q^2},\; \psi = \arctan(Q/I)
\end{equation*}

We use the Normalized Mean Square Error (NMSE) as the loss function:
\begin{equation}
\label{eq:nmse}
    L(\mathbf{h}, \hat{\mathbf{h}}) = \frac{\sum |\hat{\mathbf{h}}-\mathbf{h}|^2}{\sum |\mathbf{h}|^2}
\end{equation}
where $\mathbf{h}$ is the ground truth channel, and $\hat{\mathbf{h}}$ is the predicted channel. Unlike NeRF, we do not use Mean Square Error (MSE) as the loss function since wireless channels are typically at the scale of $10^{-1}\sim 10^{-5}$ and vary by orders of magnitude in the dataset. NMSE is more robust to changes in scale. 

Finally, it worth mentioning that similar to other NeRF work, we noticed the existence of floaters, i.e. spurious particles that "float" in space, in the trained model. We find that casting a small number (5$\sim$10) of extra rays toward random directions during the training process can effectively regulate the model to remove the floaters, thus resulting in a clean geometry of the scene.
Further details about our training pipeline and model architecture can be found in Appendix~\ref{apdx:train_details}.

\subsection{\name's Scene Representation}
\label{sec:wireless_scenes}
\begin{figure}[t]
    \centering
    \includegraphics[width=0.8\columnwidth]{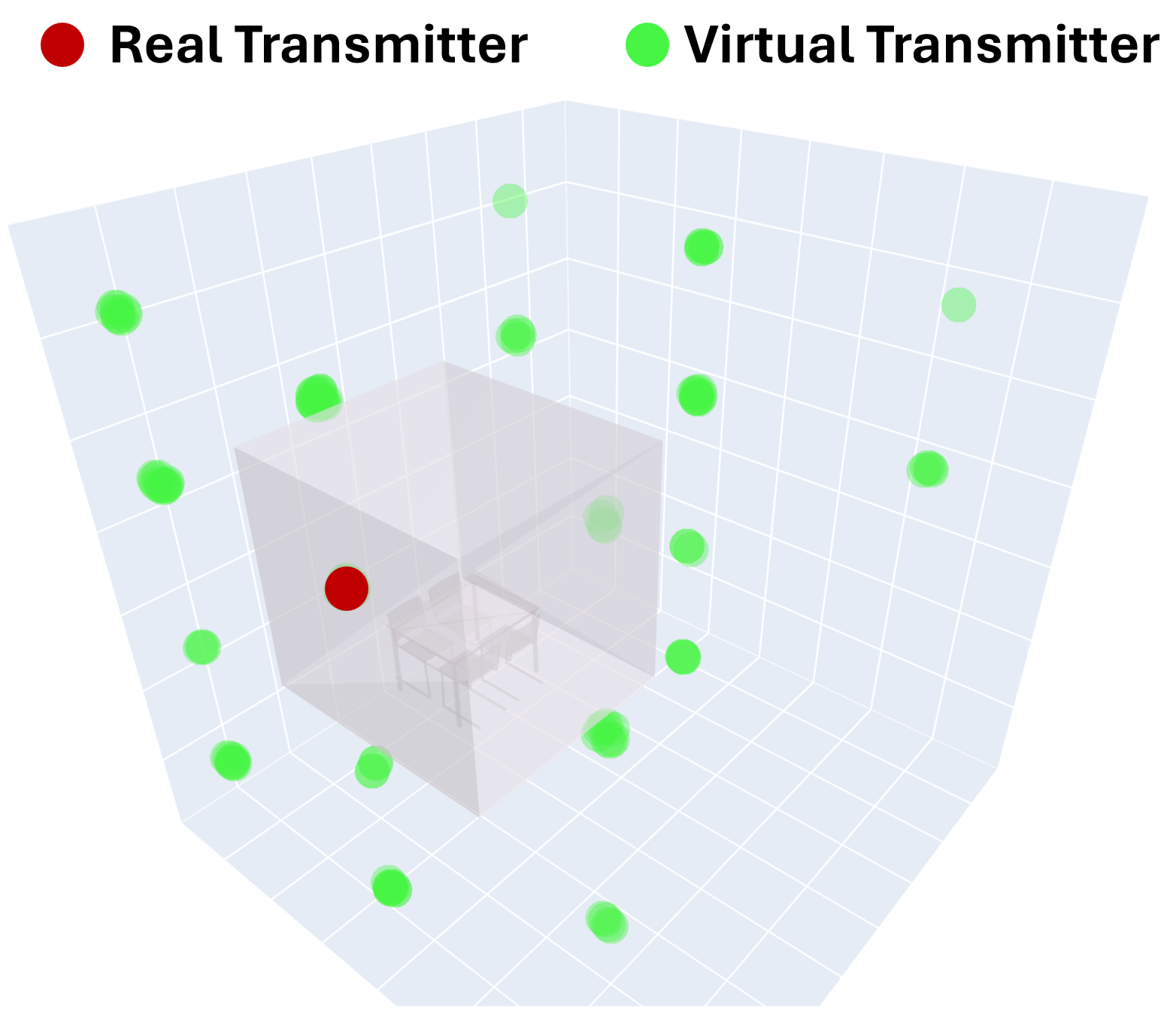}
    \caption{\textbf{A Learnt Wireless Scene:} Spots with high volume density values indicate the locations of (virtual) transmitters. The gray cube shows the 3D environment model presented in Figure~\ref{fig:conference_room}.}
    \label{fig:wireless_scene_real}
\end{figure}
\begin{figure}[t]
    \centering
    \includegraphics[width=0.8\columnwidth]{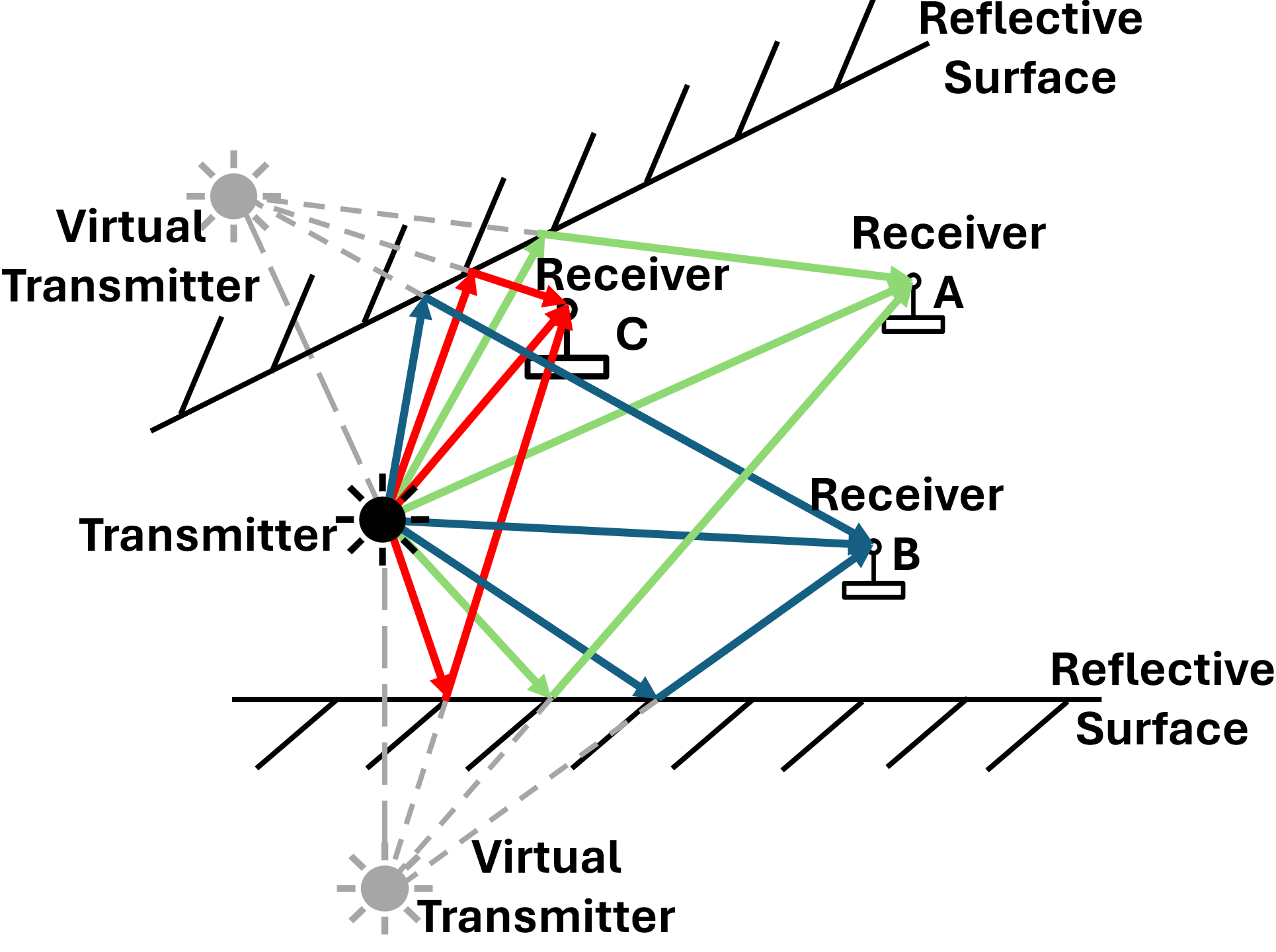}
    \caption{\textbf{The nature of wireless scenes:} While backtracing direct rays leads to the transmitter, backtracing reflected rays consistently intersect at the virtual images of the transmitter.}
    \label{fig:wireless_scene}
    \vspace{-10pt}
\end{figure}

We train the model as described in the previous section and find it has good generalization performance. However, we find that the model learns a representation of the wireless radiation scene that was different from what we originally expected. Despite that, the scene still allows the model to predict wireless fields (channels) even in room-scale environments with merely a sparse set of channel measurements. In this section, we present our findings and insights into it. 

As mentioned in Section~\ref{sec:prop_model}, we model the reflection of wireless signals as a virtual transmitter emitting signals at the reflection point on the surface. Therefore, we expect our model to learn a representation where points with high volume densities appear on the reflector surfaces of the environment. Eventually, this would allow us to reconstruct a 3D model of the environment. However, we find that our model, instead, learns to assign high volume densities at the virtual images of the real transmitter against reflective surfaces, as shown in Figure~\ref{fig:wireless_scene_real}. In this figure, we query the trained MLP model with a fine-resolution grid of 3D locations for the volume density predictions and plot the points with significant volume densities. The red dot indicates the volume blocks that match the real transmitter location. The green dots indicate the locations of virtual transmitters, which appear at virtual images of the real transmitter against each reflective surface, such as walls and tables. 

To better understand why this is the case, we illustrate a simple wireless scene with a single transmitter and two reflective surfaces in Figure~\ref{fig:wireless_scene}. Three receivers are placed in random locations to measure the wireless channel. The signals emitted by the transmitter propagate via three major paths to reach each receiver: one direct line-of-sight path and two reflection paths~\footnote{We omit the second order reflection here for simplicity; however it can also be characterized in the same manner.}. Since the receivers have no notion about the environment, they simply trace the rays back in the direction from which they arrived. Because of the law of reflection, the rays traced by all the receivers intersect at the locations of the virtual images of the transmitter and the transmitter itself. As a result, these points are sampled more frequently than other points in the space, giving the model an inductive bias to put higher volume densities at these locations. 

We note that this representation is one of many possible solutions to this optimization problem that fits the training set. However, the model seems to have such a bias that it is more likely to converge to this representation of the scene. Although this representation is not what we had initially expected, it does reveal the simple nature of radiation scenes and allow for accurate channel synthesis from sparse measurements. 
%We believe this is a very interesting property of the \name model that calls for additional exploration.

\subsection{Predicting Channels at Unvisited Locations}
\label{sec:ray_searching}
While DoA information is known for our training data, it is not for the test data. Because of this, a new ray-casting technique was required for inference time predictions. Simply casting rays across a grid of all directions is problematic. Depending on the resolution of angles used for sampling we could under or overestimate the contribution of a virtual transmitter on a channel. When the resolution is too low, virtual transmitters could be missed and when it is too high, virtual transmitters could be sampled multiple times and contribute too much to the channel. For this reason, we developed a novel ray-searching algorithm that leverages our discovery of the nature of wireless scenes to extract DoA information and thereby intelligently cast rays at new locations.

Our ray-searching algorithm consists of five steps:
\vspace{-10pt}
\begin{enumerate}
\itemsep0em 
    \item \textbf{Intersection Point Identification:} For each receiver location, $\mathbf{m}$, in the training set, perform ray marching along each DoA to identify intersection points, forming a set of potential (virtual) transmitters, $\mathbb{O}$.
    
    \item \textbf{Refine (Virtual) Transmitters:} Apply DBSCAN~\cite{ester1996density} to cluster the estimated positions in $\mathbb{O}$. Add the centroid of each cluster to a reduced set, $\mathbb{V}$.

    \item \textbf{(Virtual) Transmitter Assignment:} For each receiver location, $\mathbf{m}$, in the training set, find the closest (virtual) transmitter $\mathbf{v}\in \mathbb{V}$ along each DoA and add them to a set $\mathbb{V}_m$ of pertinent virtual transmitters for $\mathbf{m}$.

    \item \textbf{(Virtual) Transmitter Count Estimation for Testing Locations} Train a fully connected neural network $g: \mathbf{m} \rightarrow \bigm|\mathbb{V}_\mathbf{m}\bigm|$ to map receiver locations to the corresponding number of (virtual) transmitters. Then use this model to predict the size of $\mathbb{V}_\mathbf{t}$, the set of pertinent (virtual) transmitters for test location $\mathbf{t}$, for all test locations.
    
    \item \textbf{Voting-based (Virtual) Transmitter Selection:} For each $\mathbf{t}$, select its neighboring receiver locations from the training set (we use 6 nearest neighbors in practice). Populate $\mathbb{V}_t$ with the (virtual) transmitters most frequently assigned to the selected train locations. Finally, DoA can be calculated using the position of $\mathbf{t}$ and the estimated positions of (virtual) transmitters in $\mathbb{V}_t$. 
\end{enumerate}
\vspace{-10pt}

We note that although this ray-searching algorithm is able to identify the locations of (virtual) transmitters, it cannot replace the role of training a \name model for the scene, as we still need the prediction of signal amplitude and phase for each virtual transmitter.

\section{Evaluation}
\label{sec:eval}
\begin{figure}[t]
    \centering
    \includegraphics[width=\columnwidth]{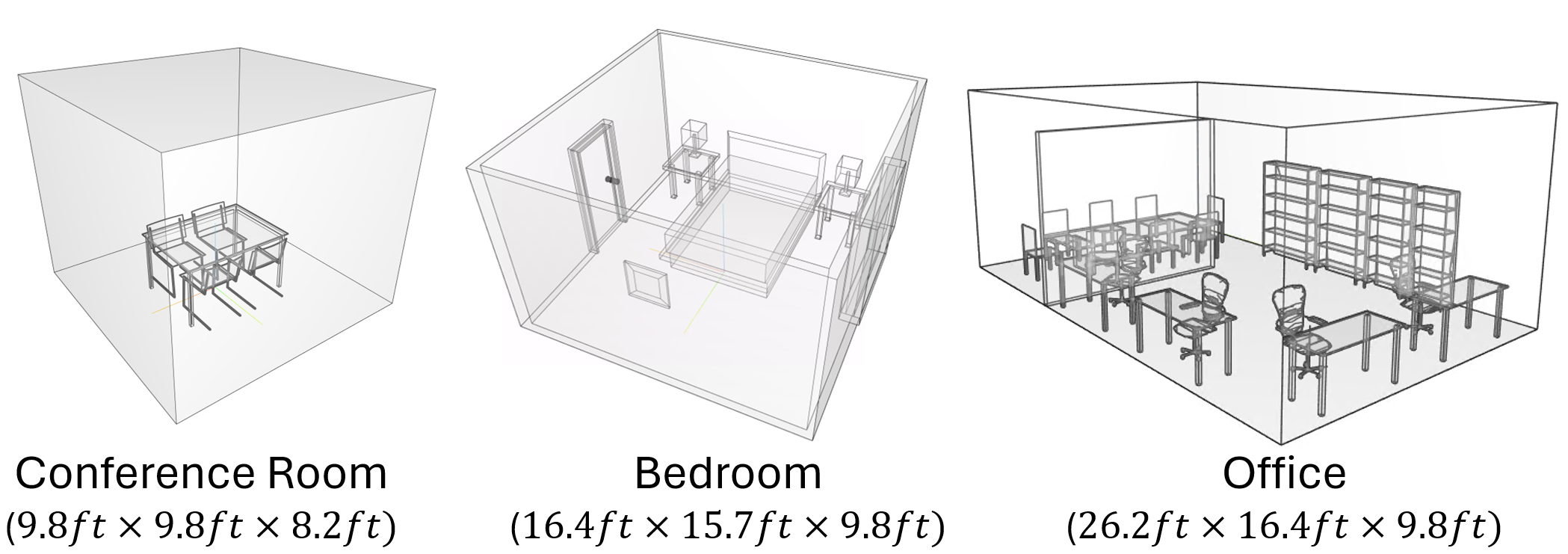}
    \vspace{-3mm}    
    \caption{3D Environment Models for simulation}
    \label{fig:env_models}
\end{figure}

To better understand the performance of \name in representing wireless scene versus its performance in predicting channels in unseen locations, we first evaluate how accurately \name can predict channels, assuming DoA is known for each receiver location. We then evaluate the performance of \name in channel prediction while the DoA is not known and we use our ray-searching algorithm (presented in \ref{sec:ray_searching}) to discover DoA of unseen locations.

\noindent\textbf{Datasets:} We generate simulation datasets for evaluation using MATLAB. We use the Image Method~\cite{yun2015ray} for ray tracing. We create datasets for three 3D environments models~\cite{matlab_conference, matlab_office, bedroom_model} of different complexities.
Figure~\ref{fig:env_models} shows the geometry and dimensions of the three models.
We put a transmitter in each environment and randomly spread 443, 975, and 1907 receivers in each environment, respectively, to collect channel measurements. We use 80\% of the measurements for training the model and the remaining 20\% for testing.

\begin{figure*}[ht]
     \centering
     \begin{subfigure}{0.33\linewidth}
        \centering
        \includegraphics[width=\linewidth]{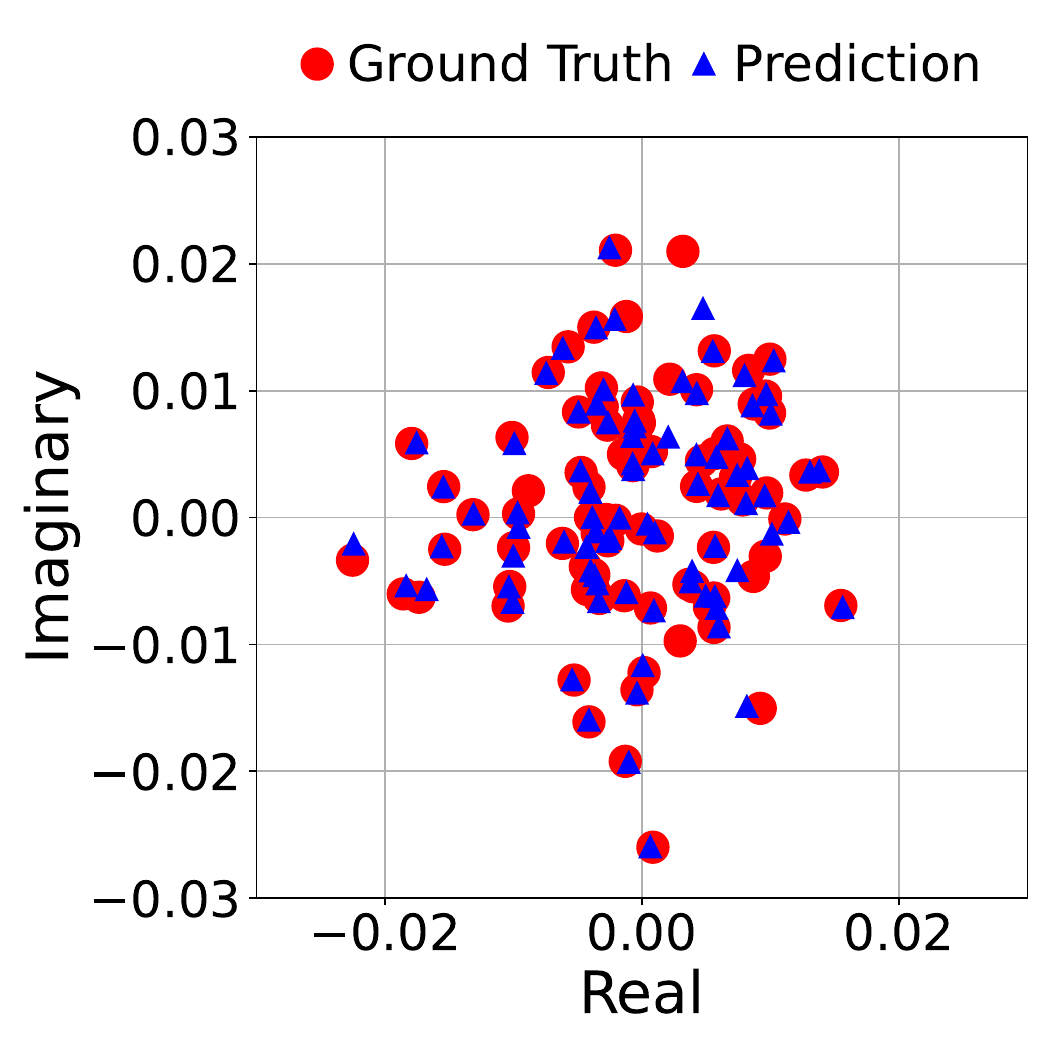}
         \caption{}
         \label{fig:channel_constellation_test}
     \end{subfigure}
     \hfill
     \begin{subfigure}{0.33\linewidth}
         \centering
        \includegraphics[width=\linewidth]{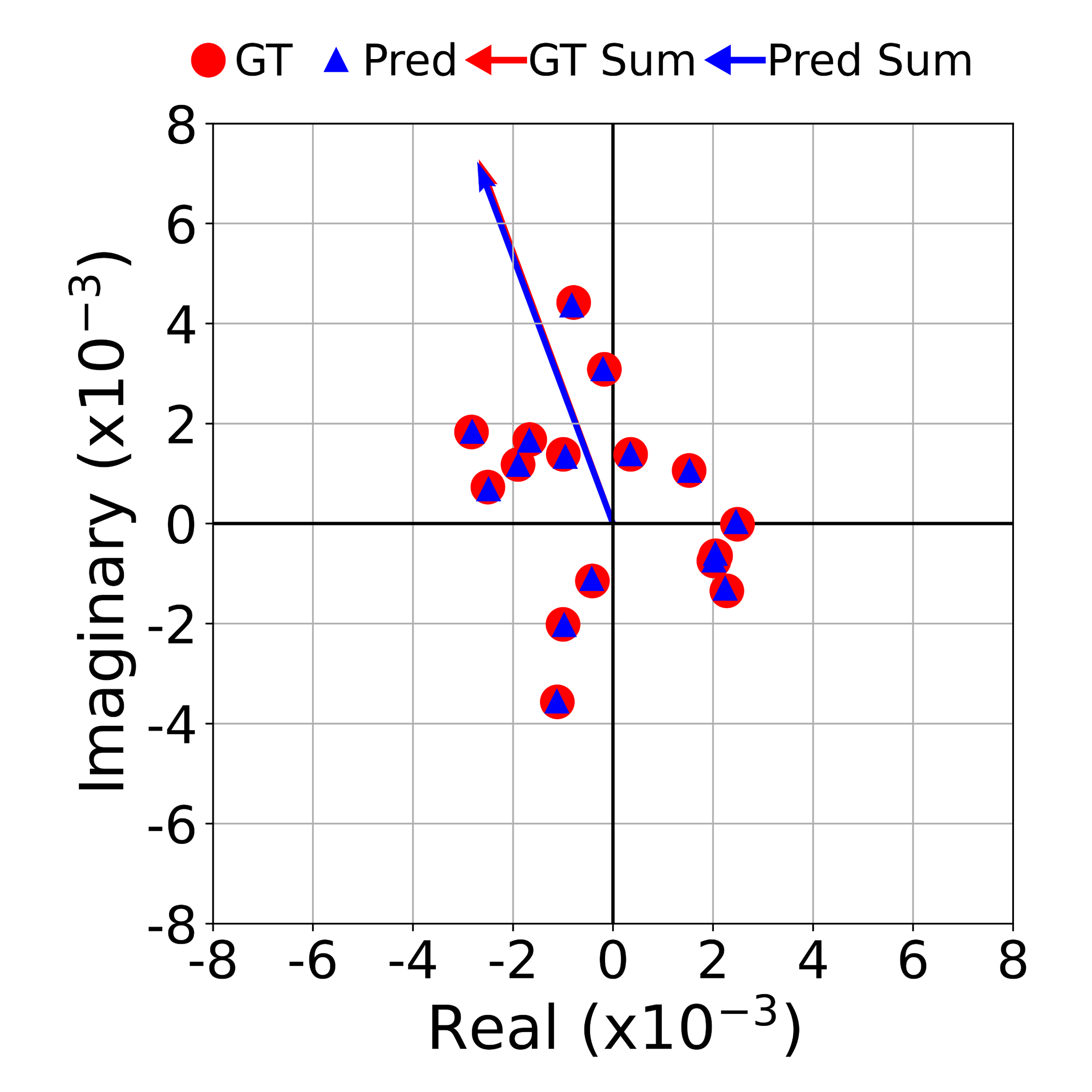}
        \caption{}
        \label{fig:single_location_channel}
     \end{subfigure}
     \hfill
    \begin{subfigure}{0.33\linewidth}
         \centering
        \includegraphics[width=\linewidth]{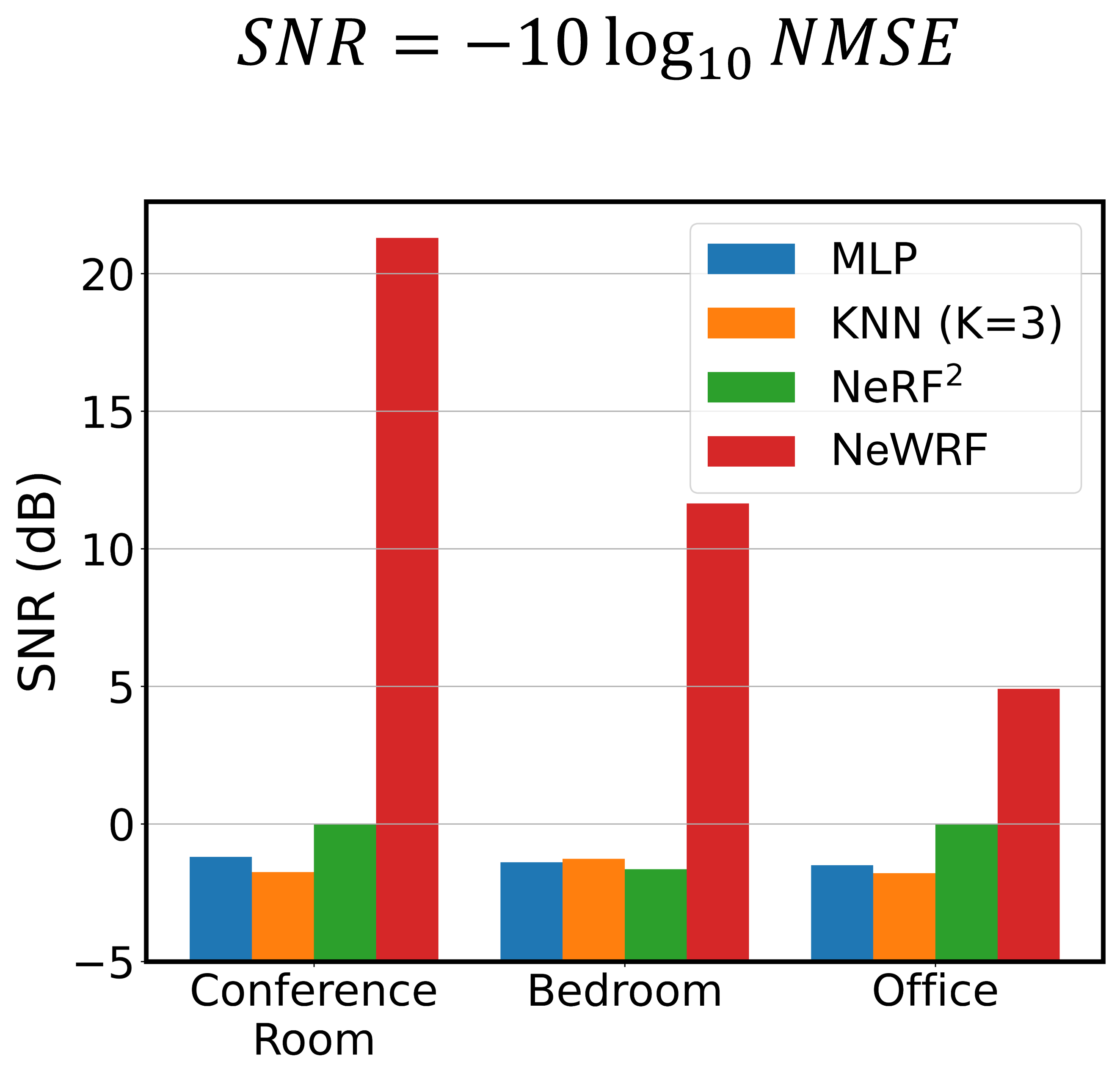}
        \caption{}
        \label{fig:baseline}
     \end{subfigure}

    % Second row

     \begin{subfigure}{0.33\linewidth}
        \centering
        \includegraphics[width=\linewidth]{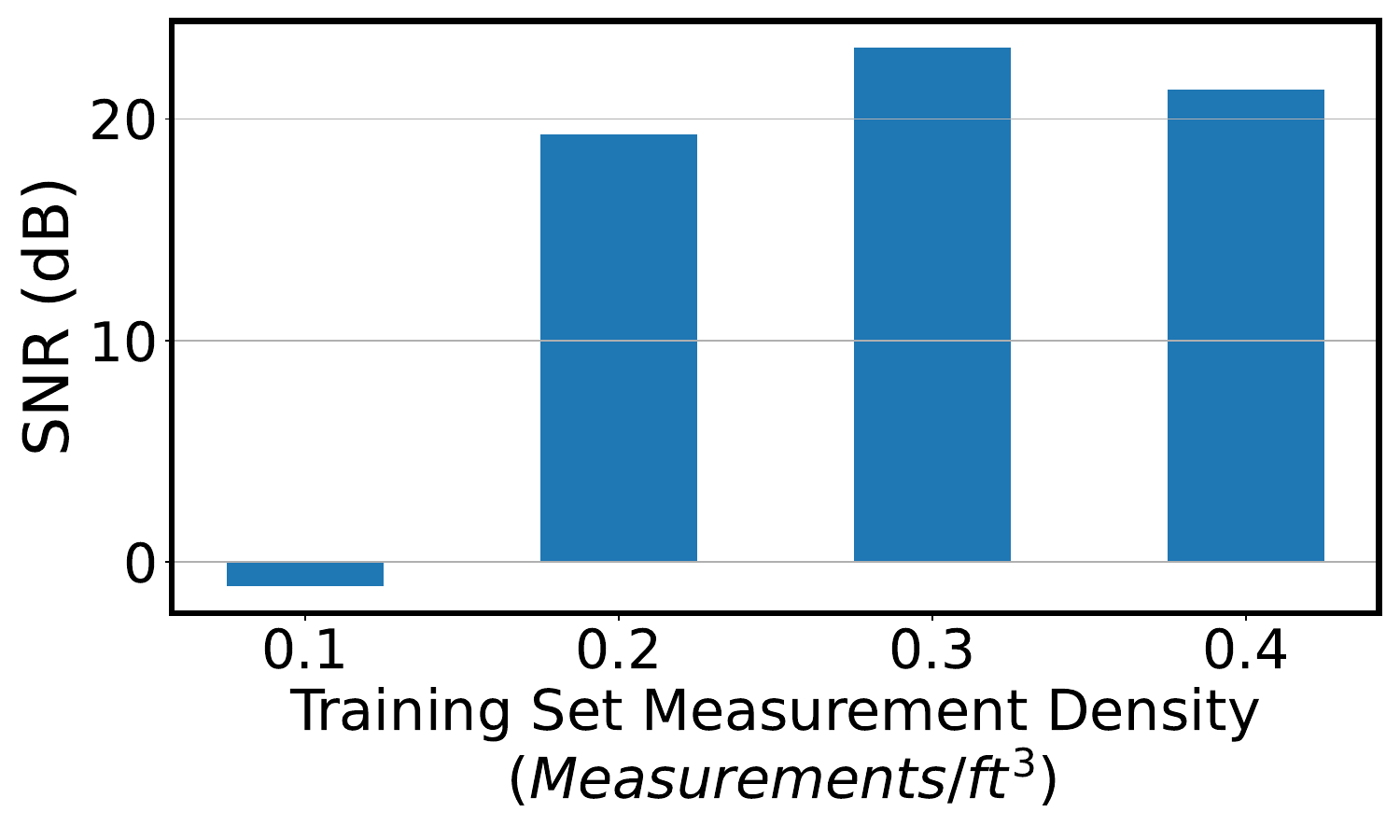}
         \caption{}
         \label{fig:trainset_percentage}
     \end{subfigure}
     \hfill
     \begin{subfigure}{0.33\linewidth}
         \centering
        \includegraphics[width=\linewidth]{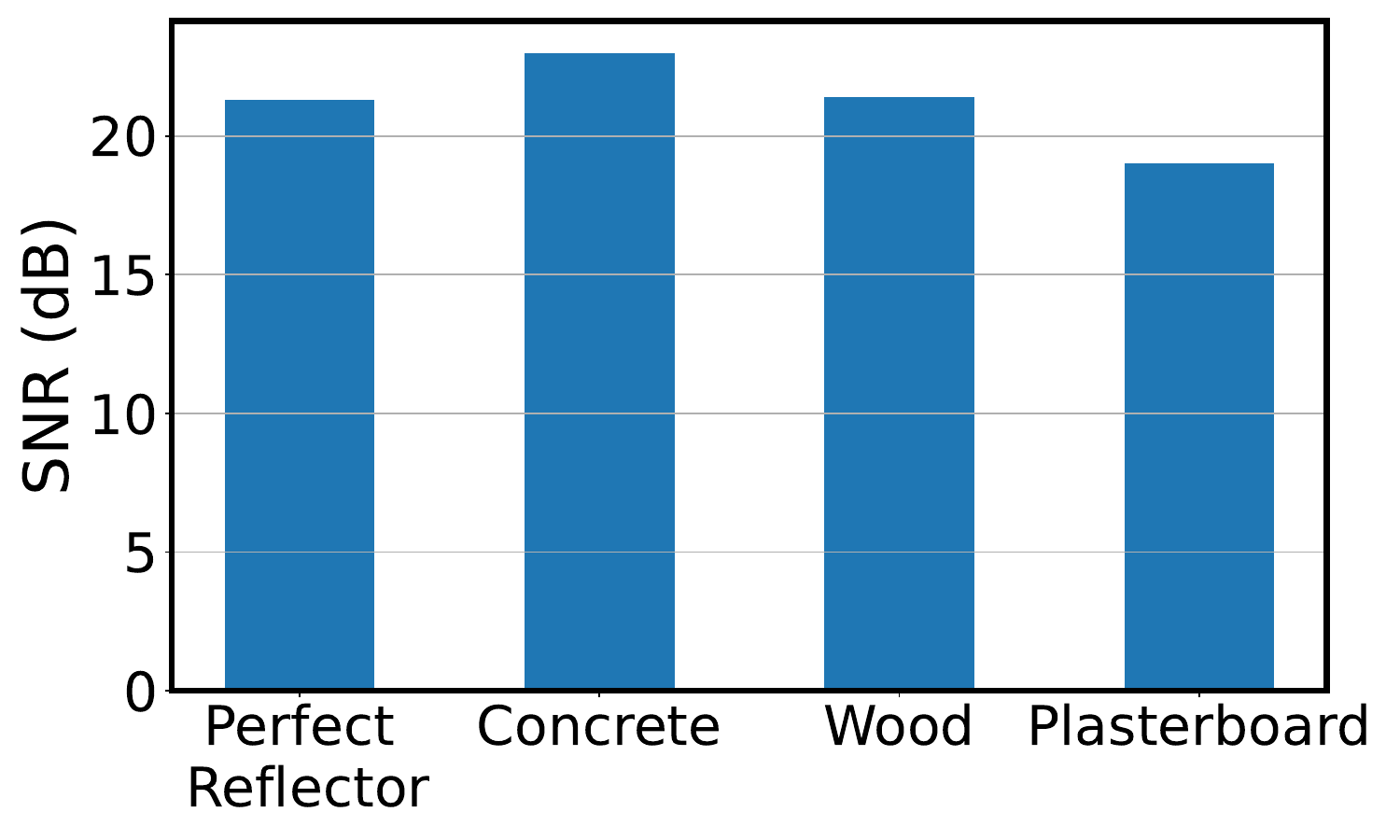}
        \caption{}
        \label{fig:material}
     \end{subfigure}
        \hfill
    \begin{subfigure}{0.33\linewidth}
         \centering
        \includegraphics[width=\linewidth]{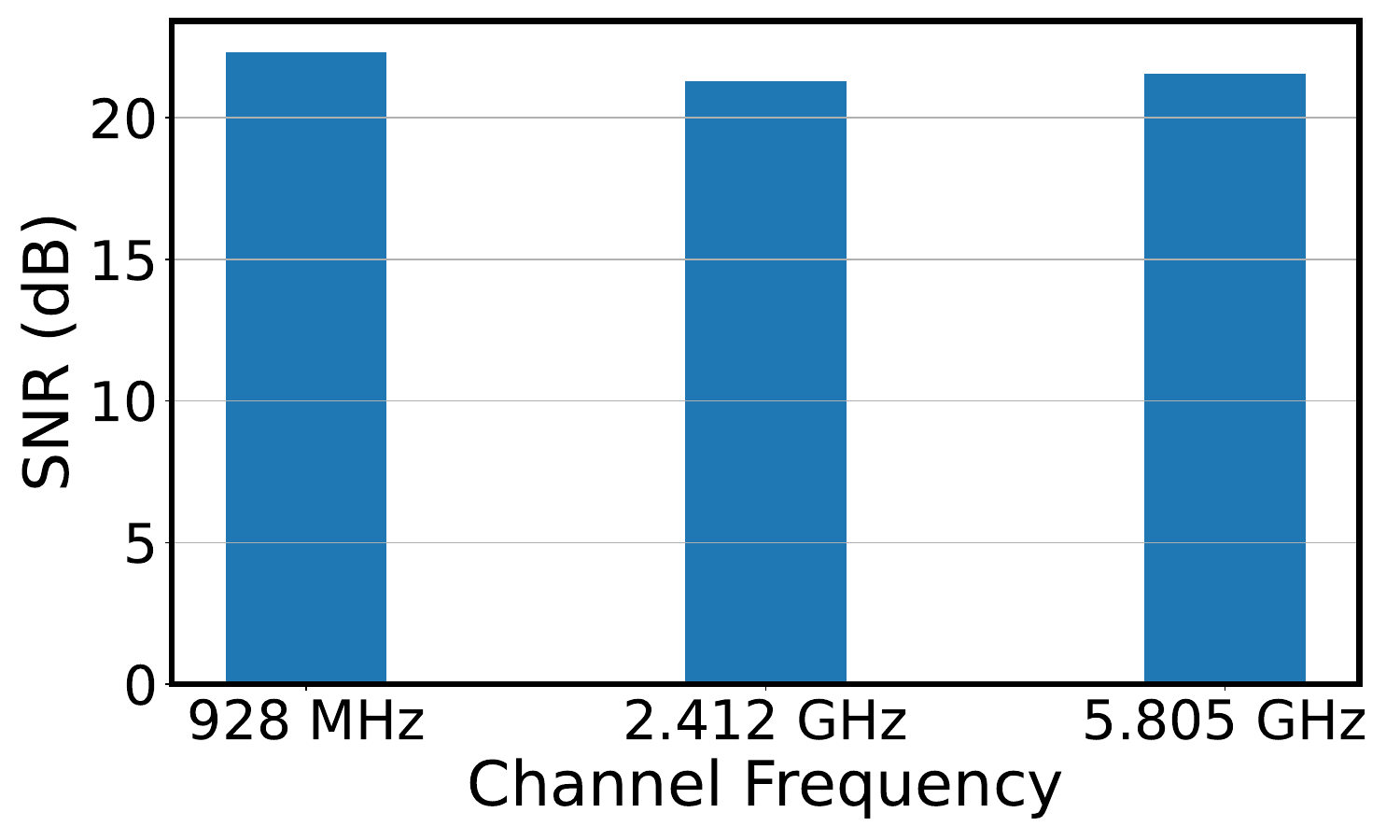}
        \caption{}
        \label{fig:freq_chs}
     \end{subfigure}
     \vspace{-5mm}
     % add psnr equation in the caption.
        \caption{\textbf{Channel Prediction Results:} a) Prediction results for conference room environment. Each dot represents the channel at one test location. b) Predicted channel components (paths) for a single location. Each dot represents a single propagation path. The arrow exhibits the sum of all paths; c) The channel prediction accuracy in different environments compared with baseline methods. d), e), and f) channel prediction accuracy for different training set sample densities, environment materials, and signal frequencies.}
        \label{fig:visual_results}
        \vspace{-10pt}
\end{figure*}
\vspace{-5pt}
\subsection{Channel Prediction Results}
We first evaluate our algorithm using our conference room dataset simulated with perfect reflector materials at a signal frequency of 2.412~GHz. We will discuss the impact of material types and signal frequencies in Section~\ref{sec:materials_freq}. 

Figure~\ref{fig:channel_constellation_test} provides a visualization of the predicted channel for the test set measurements. Since the wireless channels are represented as complex numbers, we plot the real and imaginary parts of the channels in the x and y-axis, respectively. Each dot in the plot represents the channel for a particular receiver location. As is shown, our predictions align very well with the ground truth. Moreover, although we supervise the model with merely the sum of all multipath components, \name can accurately predict individual path components for each particular receiver location as shown in Figure~\ref{fig:single_location_channel}. In this plot, each dot represents one propagation path of the wireless signal; the arrow indicates the sum of all the components. Note that this capability of predicting individual path components is very important for optimizing the configurations of base stations and relays to enhance the network coverage in practice~\cite{karanam2022foundation}. 

Next, we compare \name with three baseline methods. 

\textbf{1) K-Nearest Neighbors (KNN):} We predict the wireless channel at a test location as the average channel of the closest K locations in the training set. We pick K=3 since that gives us the lowest error.

\textbf{2) Multi-Layer Perceptrons (MLP):} We train an MLP model to directly map from the 3D location to the channel. We use an MLP with 7 linear layers, each of width 128. We train the model with the same condition as \name, as described in Appendix~\ref{apdx:train_details}.

\textbf{3) NeRF\textsuperscript{2}:} We use the open-sourced code of NeRF\textsuperscript{2}~\cite{nerf2} and train it on our datasets. Since their code assumes a dynamic transmitter scenario, we generate the counterpart simulation datasets by keeping the locations of each device the same as our datasets while swapping the roles of transmitters and receivers. 

We use Signal-to-Noise Ratio (SNR), as our evaluation metric, which is defined as the SNR = $-10\log_{10} (\text{NMSE})$, where NMSE is Normalised Mean Square Error and defined in Eq.~\eqref{eq:nmse}. Higher SNR indicates a higher prediction accuracy. For example, a 20~dB SNR is equivalent to a 1\% relative squared error in the prediction.
Figure~\ref{fig:baseline} shows the results for three environments. \name outperforms all of the baseline methods in all environments. Note, NeRF\textsuperscript{2} does not perform well on our datasets since the measurement density (0.4 $measurements/ft^3$) of our datasets is much lower than what they require (178 $measurement/ft^3$). Finally, our results show that although the performance of \name degrades as the environment becomes larger and more complex (such as a large office space with lots of furniture), it still achieves more than 5 dB SNR which is sufficient for many wireless applications.

\vspace{-5pt}
\subsection{Measurement Density of Training Set}
We further investigate the effect of training set sample density on the performance of \name. To demonstrate this, we use the conference room dataset. In this experiment, we use 25\%, 50\%, 75\%, and 100\% of the training set measurements to train our model, which corresponds to 0.1, 0.2, 0.3, and 0.4 $measurements/ft^3$, respectively. We evaluate the trained model on the same test set for a fair comparison. 
As is shown in Figure~\ref{fig:trainset_percentage}, \name can effectively learn the wireless radiation scene, and predict the channel accurately with sample density as low as 0.2 $measurements/ft^3$, which is around 800$\times$ lower than that of NeRF\textsuperscript{2}\cite{nerf2}. This would significantly reduce the cost of wireless site surveys and improve the capability of dead spot detection. 

\subsection{Impact of Materials Types and Signal Frequencies}
\label{sec:materials_freq}
The attenuation and phase rotation of wireless signals during reflection are largely determined by the material that constitutes the reflector. In particular, the material's electrical properties, such as permittivity and conductivity, play significant roles. As such, the wireless properties of an environment can change greatly as the materials it is made of change. Furthermore, even the same materials have different behaviors when they interact with signals of different frequencies. \name is able to characterize both of these effects, as shown in Figure~\ref{fig:material} and~\ref{fig:freq_chs}, where we train a separate model for each material type and signal frequency. \name maintains high prediction accuracy in all these scenarios.

\begin{figure}
    \centering
    \includegraphics[width=0.8\columnwidth]{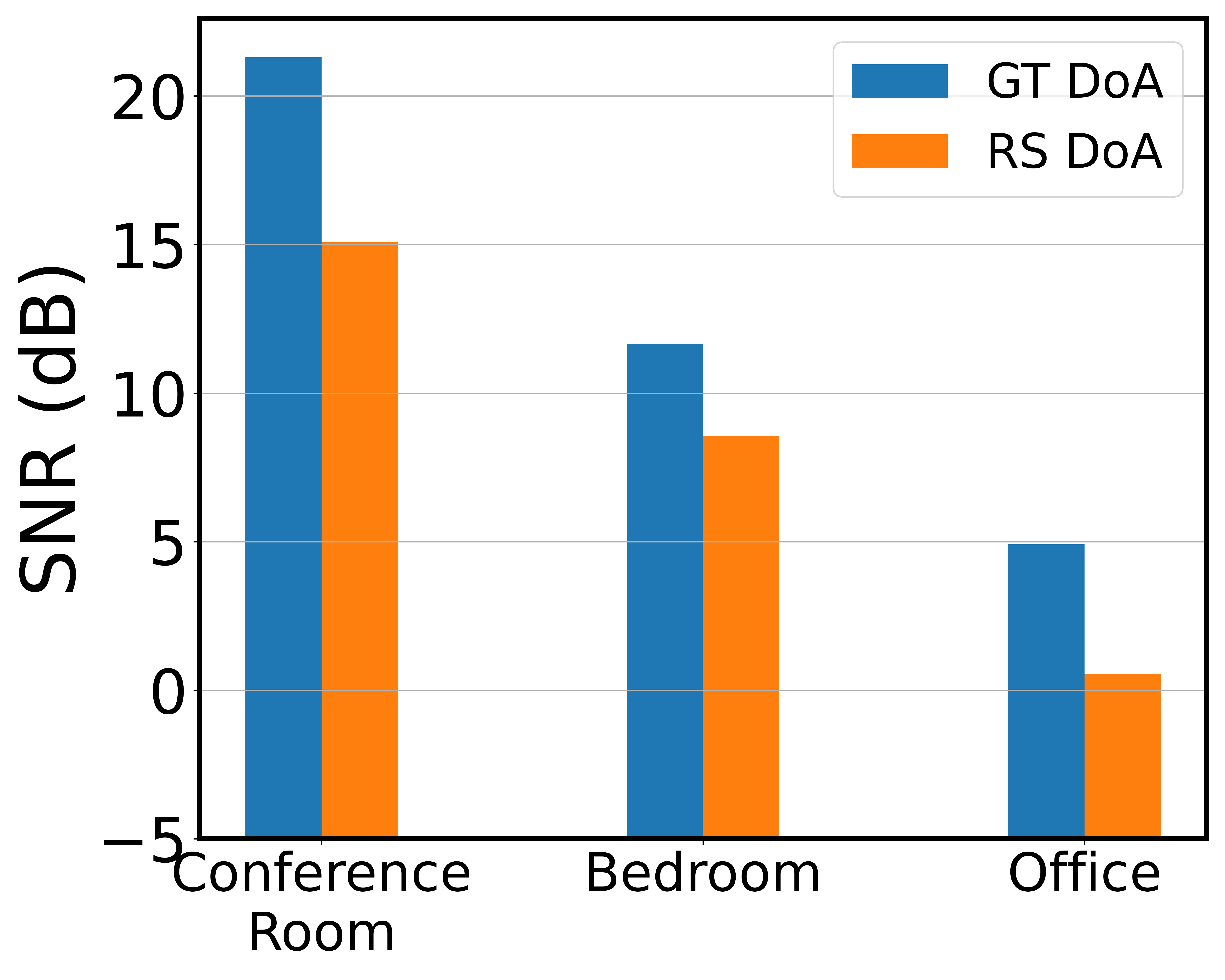}
     \caption{\textbf{Ray-searching algorithm performance:} Performance of \name in channel prediction when DoA is known using ground truth or discovered using our Ray-searching algorithm. GT stands for ground truth, and RS stands for the Ray-Searching algorithm presented in ~\ref{sec:ray_searching}.}
     \label{fig:ray_search_eval}
     \vspace{-25pt}
\end{figure}
     
% Next, we are interested in seeing the robustness of \name against the influence of noise in wireless measurements. In this experiment, we add artificial Gaussian white noise to the simulated channels. The ratio between the channel power and added noise power is denoted as channel SNR. Figure~\ref{fig:ch_noise} shows the results, the blue line shows the results of adding a fixed noise, i.e. bias, to the ground truth channel in the training set. As is expected, the prediction accuracy reduces as the channel SNR decreases. However, typical wireless systems today achieve more than 20 dB channel SNR in practice for reliable communication, where \name performs well in such SNRs. Also, we note that collecting multiple measurements with different biases can improve the robustness. The orange line in Figure~\ref{fig:ch_noise} demonstrates this effect, where we use measurements with different additive noises in each iteration of training. The model is able to maintain good performance even at very low SNRs. This is because the model implicitly averages these measurements to reduce the bias during the training. 

\subsection{Ray-searching Algorithm Performance} 
So far we evaluated the performance of \name while DoA is known. Here, we evaluate the performance of \name when we use our ray-searching algorithm to discover the DoA for each location. Figure~\ref{fig:ray_search_eval} shows the results. Our results show that due to the error in DoA estimation, the performance of \name in channel estimation degrades by a small margin (3$\sim$5 dB) when we use the ray-searching algorithm. This error is mainly due to a slight discrepancy of DoA between the test location and its nearest neighbors. Although this SNR loss is negligible, one can improve it by using slightly higher measurement density.

%However, this can be improved by using a slightly denser measurement. For example, when we increase the density to 4.0 measurements/$ft^3$, \name achieves SNR of 18 dB.

\section{Limitations and Future work}
\label{sec:disc}
In this work, we bring NeRF to radio frequency and demonstrate the promise of reconstructing wireless radiation scenes to facilitate site surveys. However, there remains a vast range of directions to investigate to unleash the power of NeRF for wireless applications. Here we highlight a few of them to help stimulate research in these directions:

a) Relaxing phase requirement: accurate phase measurement is challenging in real communication systems due to the synchronization error between the transmitter and receiver. The presented \name algorithm, which relies on narrowband channels for training, does not perform well if the phase is inaccurate or not provided. One potential solution might be using wideband channel measurements for training since wideband channels translate phase into the variation of amplitude across different frequencies, which can be measured accurately.

b) Relaxing DoA requirement: accurate DoA measurement is required for ray-casting in the current \name algorithm, the error tolerance is within one degree based on our experiments. In practice, achieving such a measurement resolution would require a very large antenna array. Further studies are needed to relax this constraint.

c) Extending to larger environments: although \name outperforms all the baselines for the three environments studied, its performance reduction with the increasing environment complexity is observed. Scaling to more complex environments requires further research. One potential solution would be fusing visual and wireless radiation fields, where visual images provide more information about the environment to facilitate the learning of wireless fields.

\vspace{-5pt}
\section{Conclusion}
In this work, we propose \name, the first NeRF-based learning framework for wireless channel prediction with sparse measurements. We solve a series of challenges to enable the neural synthesis of wireless channels and, for the first time, reveal the nature of wireless scenes. We believe \name represents the first step towards the practical use of NeRF in future wireless applications.

\section*{Acknowledgements}
We thank the UCLA ICON group and the reviewers for their insightful comments. We also thank Yadi Cao for his enlightening discussion. This work is partially supported by UCLA, and NSF awards 2238245 and 2146492. 

\section*{Impact Statement}
This paper presents work whose goal is to advance the field of 
Machine Learning. There are many potential societal consequences 
of our work, none which we feel must be specifically highlighted here.

\nocite{langley00}

\bibliography{ref}

\begin{thebibliography}{25}
\providecommand{\natexlab}[1]{#1}
\providecommand{\url}[1]{\texttt{#1}}
\expandafter\ifx\csname urlstyle\endcsname\relax
  \providecommand{\doi}[1]{doi: #1}\else
  \providecommand{\doi}{doi: \begingroup \urlstyle{rm}\Url}\fi

\bibitem[Azinovi{\'c} et~al.(2022)Azinovi{\'c}, Martin-Brualla, Goldman, Nie{\ss}ner, and Thies]{azinovic2022neural}
Azinovi{\'c}, D., Martin-Brualla, R., Goldman, D.~B., Nie{\ss}ner, M., and Thies, J.
\newblock Neural rgb-d surface reconstruction.
\newblock In \emph{Proceedings of the IEEE/CVF Conference on Computer Vision and Pattern Recognition}, pp.\  6290--6301, 2022.

\bibitem[Ester et~al.(1996)Ester, Kriegel, Sander, Xu, et~al.]{ester1996density}
Ester, M., Kriegel, H.-P., Sander, J., Xu, X., et~al.
\newblock A density-based algorithm for discovering clusters in large spatial databases with noise.
\newblock In \emph{kdd}, volume~96, pp.\  226--231, 1996.

\bibitem[Formis et~al.(2023)Formis, Scanzio, Cena, and Valenzano]{Formis_2023}
Formis, G., Scanzio, S., Cena, G., and Valenzano, A.
\newblock Predicting wireless channel quality by means of moving averages and regression models.
\newblock In \emph{2023 IEEE 19th International Conference on Factory Communication Systems (WFCS)}. IEEE, April 2023.
\newblock \doi{10.1109/wfcs57264.2023.10144122}.
\newblock URL \url{http://dx.doi.org/10.1109/WFCS57264.2023.10144122}.

\bibitem[Foutz et~al.(2022)Foutz, Spanias, and Banavar]{foutz2022narrowband}
Foutz, J., Spanias, A., and Banavar, M.
\newblock \emph{Narrowband Direction of Arrival Estimation for Antenna Arrays}.
\newblock Synthesis Lectures on Antennas. Springer International Publishing, 2022.
\newblock ISBN 9783031015373.
\newblock URL \url{https://books.google.com/books?id=Y4RyEAAAQBAJ}.

\bibitem[Free3D(2023)]{bedroom_model}
Free3D.
\newblock Bedroom 3d model.
\newblock \url{https://free3d.com/3d-model/bedroom-86855.html}, 2023.

\bibitem[Kar \& Dappuri(2018)Kar and Dappuri]{site_survey}
Kar, P. and Dappuri, B.
\newblock Site survey and radio frequency planning for the deployment of next generation wlan.
\newblock In \emph{2018 International Conference on Wireless Communications, Signal Processing and Networking (WiSPNET)}, pp.\  1--4, 2018.
\newblock \doi{10.1109/WiSPNET.2018.8538731}.

\bibitem[Karanam \& Mostofi(2023)Karanam and Mostofi]{karanam2022foundation}
Karanam, C.~R. and Mostofi, Y.
\newblock A foundation for wireless channel prediction and full ray makeup estimation using an unmanned vehicle.
\newblock \emph{IEEE Sensors Journal}, 23\penalty0 (18):\penalty0 21452--21462, 2023.
\newblock \doi{10.1109/JSEN.2023.3299951}.

\bibitem[Kingma \& Ba(2014)Kingma and Ba]{kingma2014adam}
Kingma, D.~P. and Ba, J.
\newblock Adam: A method for stochastic optimization.
\newblock \emph{arXiv preprint arXiv:1412.6980}, 2014.

\bibitem[Krijestorac et~al.(2021)Krijestorac, Hanna, and Cabric]{krijestorac2021spatial}
Krijestorac, E., Hanna, S., and Cabric, D.
\newblock Spatial signal strength prediction using 3d maps and deep learning.
\newblock In \emph{ICC 2021-IEEE international conference on communications}, pp.\  1--6. IEEE, 2021.

\bibitem[Ma et~al.(2019)Ma, Zhou, and Wang]{ma2019wifi}
Ma, Y., Zhou, G., and Wang, S.
\newblock Wifi sensing with channel state information: A survey.
\newblock \emph{ACM Computing Surveys (CSUR)}, 52\penalty0 (3):\penalty0 1--36, 2019.

\bibitem[Malmirchegini \& Mostofi(2012)Malmirchegini and Mostofi]{malmirchegini2012spatial}
Malmirchegini, M. and Mostofi, Y.
\newblock On the spatial predictability of communication channels.
\newblock \emph{IEEE Transactions on Wireless Communications}, 11\penalty0 (3):\penalty0 964--978, 2012.

\bibitem[Martin-Brualla et~al.(2021)Martin-Brualla, Radwan, Sajjadi, Barron, Dosovitskiy, and Duckworth]{martin2021nerf}
Martin-Brualla, R., Radwan, N., Sajjadi, M.~S., Barron, J.~T., Dosovitskiy, A., and Duckworth, D.
\newblock Nerf in the wild: Neural radiance fields for unconstrained photo collections.
\newblock In \emph{Proceedings of the IEEE/CVF Conference on Computer Vision and Pattern Recognition}, pp.\  7210--7219, 2021.

\bibitem[MATLAB(2023{\natexlab{a}})]{matlab_conference}
MATLAB.
\newblock Indoor mimo-ofdm communication link using ray tracing.
\newblock https://www.mathworks.com/help/comm/ug/indoor-mimo-ofdm-communication-link-using-ray-tracing.html, 2023{\natexlab{a}}.

\bibitem[MATLAB(2023{\natexlab{b}})]{matlab_office}
MATLAB.
\newblock Three-dimensional indoor positioning with 802.11az fingerprinting and deep learning.
\newblock https://www.mathworks.com/help/wlan/ug/three-dimensional-indoor-positioning-with-802-11az-fingerprinting-and-deep-learning.html, 2023{\natexlab{b}}.

\bibitem[Mildenhall et~al.(2021)Mildenhall, Srinivasan, Tancik, Barron, Ramamoorthi, and Ng]{mildenhall2021nerf}
Mildenhall, B., Srinivasan, P.~P., Tancik, M., Barron, J.~T., Ramamoorthi, R., and Ng, R.
\newblock Nerf: Representing scenes as neural radiance fields for view synthesis.
\newblock \emph{Communications of the ACM}, 65\penalty0 (1):\penalty0 99--106, 2021.

\bibitem[Mildenhall et~al.(2022)Mildenhall, Hedman, Martin-Brualla, Srinivasan, and Barron]{mildenhall2022nerf}
Mildenhall, B., Hedman, P., Martin-Brualla, R., Srinivasan, P.~P., and Barron, J.~T.
\newblock Nerf in the dark: High dynamic range view synthesis from noisy raw images.
\newblock In \emph{Proceedings of the IEEE/CVF Conference on Computer Vision and Pattern Recognition}, pp.\  16190--16199, 2022.

\bibitem[Mittra(2016)]{mittra2016computational}
Mittra, R.
\newblock \emph{Computational electromagnetics}.
\newblock Springer, 2016.

\bibitem[Molisch(2012)]{molisch2012wireless}
Molisch, A.~F.
\newblock \emph{Wireless communications}.
\newblock John Wiley \& Sons, 2012.

\bibitem[Orekondy et~al.(2022)Orekondy, Kumar, Kadambi, Ye, Soriaga, and Behboodi]{orekondy2022winert}
Orekondy, T., Kumar, P., Kadambi, S., Ye, H., Soriaga, J., and Behboodi, A.
\newblock Winert: Towards neural ray tracing for wireless channel modelling and differentiable simulations.
\newblock In \emph{The Eleventh International Conference on Learning Representations}, 2022.

\bibitem[Pumarola et~al.(2021)Pumarola, Corona, Pons-Moll, and Moreno-Noguer]{pumarola2021d}
Pumarola, A., Corona, E., Pons-Moll, G., and Moreno-Noguer, F.
\newblock D-nerf: Neural radiance fields for dynamic scenes.
\newblock In \emph{Proceedings of the IEEE/CVF Conference on Computer Vision and Pattern Recognition}, pp.\  10318--10327, 2021.

\bibitem[Remcom(2023)]{remcom_wirelessinsite}
Remcom.
\newblock Wireless insite.
\newblock \url{https://www.remcom.com/wireless-insite-em-propagation-software}, 2023.
\newblock Version 3.4.4.

\bibitem[Roessle et~al.(2022)Roessle, Barron, Mildenhall, Srinivasan, and Nie{\ss}ner]{roessle2022dense}
Roessle, B., Barron, J.~T., Mildenhall, B., Srinivasan, P.~P., and Nie{\ss}ner, M.
\newblock Dense depth priors for neural radiance fields from sparse input views.
\newblock In \emph{Proceedings of the IEEE/CVF Conference on Computer Vision and Pattern Recognition}, pp.\  12892--12901, 2022.

\bibitem[Varshney et~al.(2023)Varshney, Gangal, Sharique, and Ansari]{Rajat2023}
Varshney, R., Gangal, C., Sharique, M., and Ansari, M.~S.
\newblock Deep learning based wireless channel prediction: 5g scenario.
\newblock \emph{Procedia Computer Science}, 218:\penalty0 2626--2635, 2023.
\newblock ISSN 1877-0509.
\newblock \doi{https://doi.org/10.1016/j.procs.2023.01.236}.
\newblock URL \url{https://www.sciencedirect.com/science/article/pii/S1877050923002363}.
\newblock International Conference on Machine Learning and Data Engineering.

\bibitem[Yun \& Iskander(2015)Yun and Iskander]{yun2015ray}
Yun, Z. and Iskander, M.~F.
\newblock Ray tracing for radio propagation modeling: Principles and applications.
\newblock \emph{IEEE access}, 3:\penalty0 1089--1100, 2015.

\bibitem[Zhao et~al.(2023)Zhao, An, Pan, and Yang]{nerf2}
Zhao, X., An, Z., Pan, Q., and Yang, L.
\newblock \emph{NeRF2: Neural Radio-Frequency Radiance Fields}.
\newblock Association for Computing Machinery, New York, NY, USA, 2023.
\newblock ISBN 9781450399906.
\newblock URL \url{https://doi.org/10.1145/3570361.3592527}.

\end{thebibliography}
\bibliographystyle{icml2024}

%%%%%%%%%%%%%%%%%%%%%%%%%%%%%%%%%%%%%%%%%%%%%%%%%%%%%%%%%%%%%%%%%%%%%%%%%%%%%%%
%%%%%%%%%%%%%%%%%%%%%%%%%%%%%%%%%%%%%%%%%%%%%%%%%%%%%%%%%%%%%%%%%%%%%%%%%%%%%%%
% APPENDIX
%%%%%%%%%%%%%%%%%%%%%%%%%%%%%%%%%%%%%%%%%%%%%%%%%%%%%%%%%%%%%%%%%%%%%%%%%%%%%%%
%%%%%%%%%%%%%%%%%%%%%%%%%%%%%%%%%%%%%%%%%%%%%%%%%%%%%%%%%%%%%%%%%%%%%%%%%%%%%%%
\newpage
\appendix
\onecolumn
\section{Implementation Details}
\label{apdx:train_details}
% \begin{figure*}[t]
%     \centering
%     \includegraphics[width=\textwidth]{Figures/pipeline.png}
%     \caption{End-to-End Training Pipeline}
%     \label{fig:pipeline}
% \end{figure*}

% \begin{figure*}[t]
%     \centering
%     \includegraphics[width=\textwidth]{Figures/model_architecture_V2.png}
%     \caption{Model Architecture}
%     \label{fig:arch}
% \end{figure*}

\begin{figure*}[ht]
    \centering
    \begin{minipage}{0.73\linewidth}
        \centering
        \includegraphics[width=\linewidth]{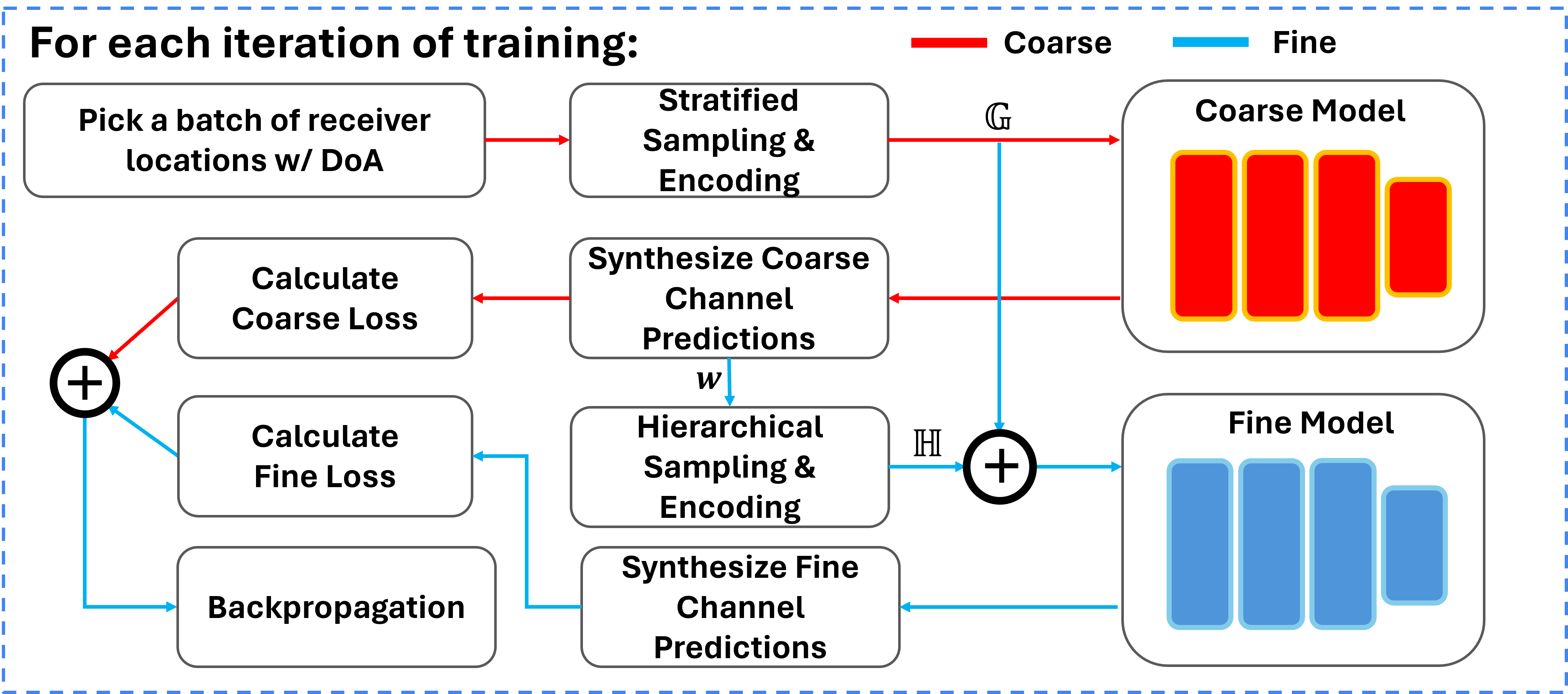} % first figure itself
        \caption{\name's Training Pipeline}
        \label{fig:pipeline}
    \end{minipage}\hfill
    \begin{minipage}{0.25\linewidth}
        \centering
        \includegraphics[width=\linewidth]{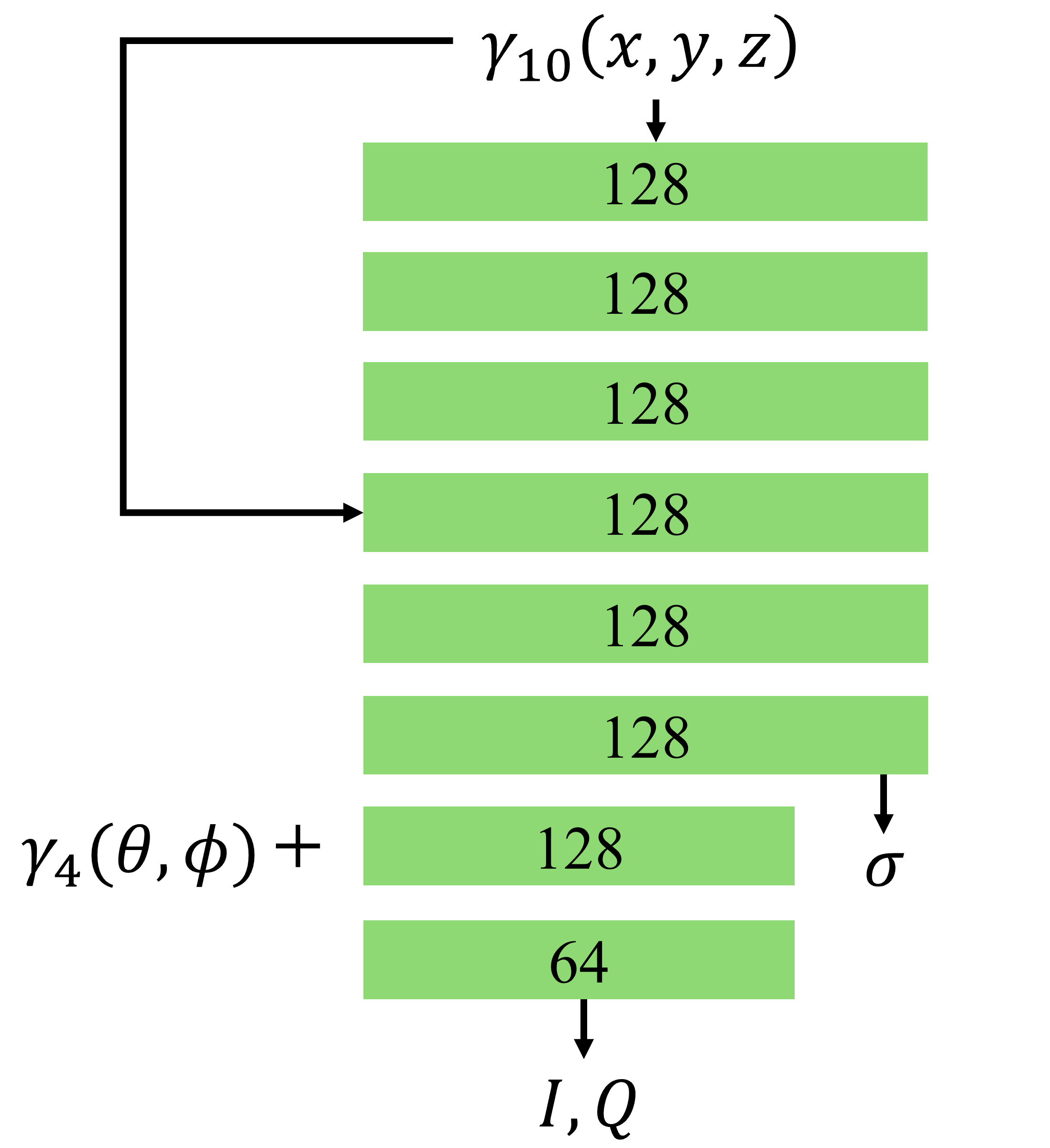} % second figure itself
        \caption{\name's MLP Architecture. Both Coarse and Fine model use the same architecture.}
        \label{fig:arch}
    \end{minipage}
\end{figure*}

We implement \name in Pytorch. Figure~\ref{fig:pipeline} shows the training pipeline. We use a batch size of 32 for each iteration of training. As in NeRF, we use positional encoding to map the input coordinates to higher dimensional space to facilitate the model to learn high-frequency variation of the wireless scene~\cite{mildenhall2021nerf}. We use 10 frequencies to encode the spatial coordinates $(x,y,z)$, and 4 frequencies for the direction coordinates. 

Figure~\ref{fig:arch} shows the MLP model architecture of \name. It consists of 7 linear layers of width 128 and one output layer of width 64. All layers as well as the $\sigma$ output are activated by ReLU functions. The I, Q outputs are activated with an $tanh$ function to keep the output value within range [-1, 1]. A single skip connection concatenates the input with the fourth linear layer. $\gamma(\cdot)$ represents the positional encoding function. 

We use the same architecture for both "coarse" and "fine" models. A weighted sum of "coarse" and "fine" model loss is used for supervision: 
\[
loss = 0.1\times coarse\; loss + 0.9\times fine\; loss
\]
We use Adam optimizer~\cite{kingma2014adam} with an initial learning rate of $5\times10^{-4}$, and a \textsc{ReduceLROnPlateau} learning rate scheduler with patience 10 and factor 0.9. The other parameters are left as default. We train our model on a single NVIDIA A100 GPU, and it typically takes $\sim$100k iterations to converge. 

The sampling ranges and resolutions along each ray are adjusted for each environment due to their different dimensions. The sampling range is set to 9~m, 15~m, and 24~m for the conference room, bedroom, and office, respectively. For the conference room and bedroom, we take 128 samples per ray for the coarse stratified sampling, and another 128 samples per ray in the hierarchical sampling to enhance the granularity. For the office environment, due to its complexity, we double the number of samples for both sampling stages.

% \section{More complex propagation effects}
% We run simulations in WirelessInsite for the same conference room environment, transmitter, and receiver configurations to study more complex propagation effects, such as scattering, penetration, and higher-order reflections on the performance of \name. The evaluation results are shown in Figure~\ref{fig:wi_vs_matlab}. MATLAB dataset accounts for up to second-order reflections (R) and first-order diffraction (D), whereas the WirelessInsite dataset includes up to third-order reflections and transmissions (T) (penetration through objects) and first-order diffraction. Note that higher-order reflections do not alter the channel much, as their amplitudes are orders lower due to the attenuation involved. we observe only a small degradation ($\sim$3~dB) in the prediction accuracy due to the extra complexity.

\begin{figure*}[ht]
    \centering
    \begin{minipage}{0.33\linewidth}
        \centering
        \includegraphics[width=\linewidth]{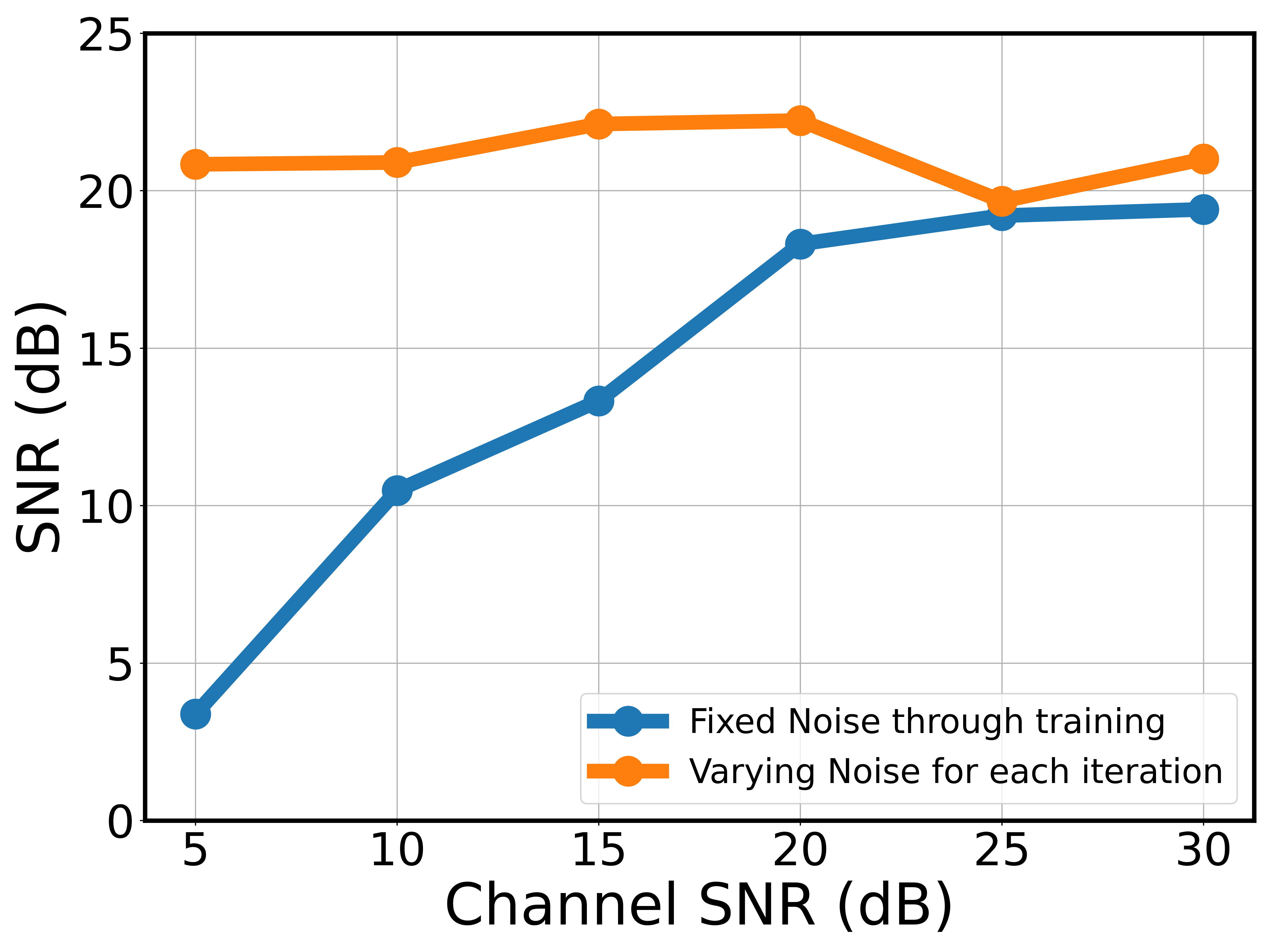} % first figure itself
        \caption{\name's performance on adding noise to the channel}
        \label{fig:ch_noise}
    \end{minipage}\hfill
    \begin{minipage}{0.66\linewidth}
        \centering
     \begin{subfigure}{0.48\linewidth}
        \centering
        \includegraphics[width=\linewidth]{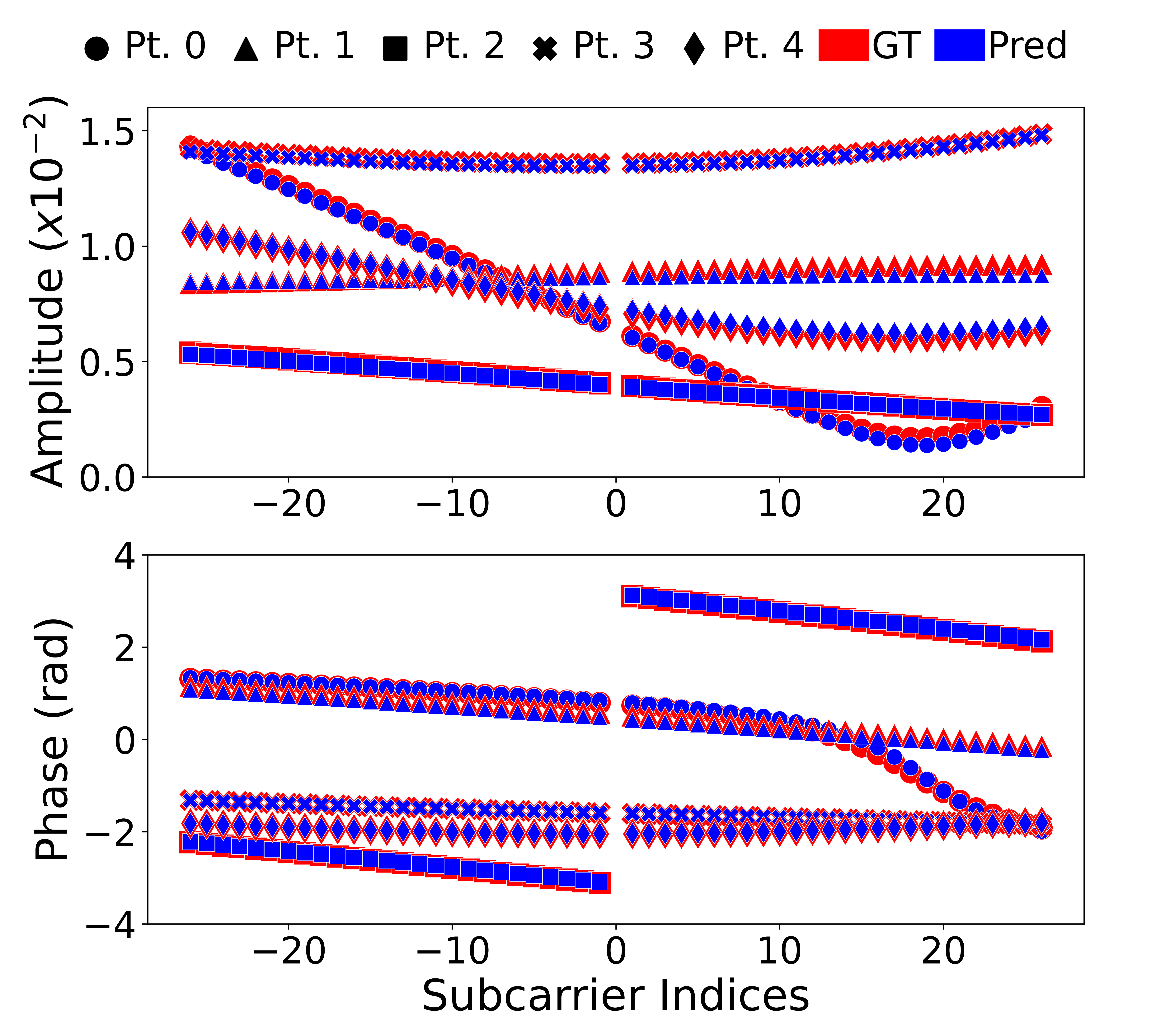}
         \caption{}
         \label{fig:subcarrier}
     \end{subfigure}
     % \hfill
    \begin{subfigure}{0.48\linewidth}
         \centering
        \includegraphics[width=\linewidth]{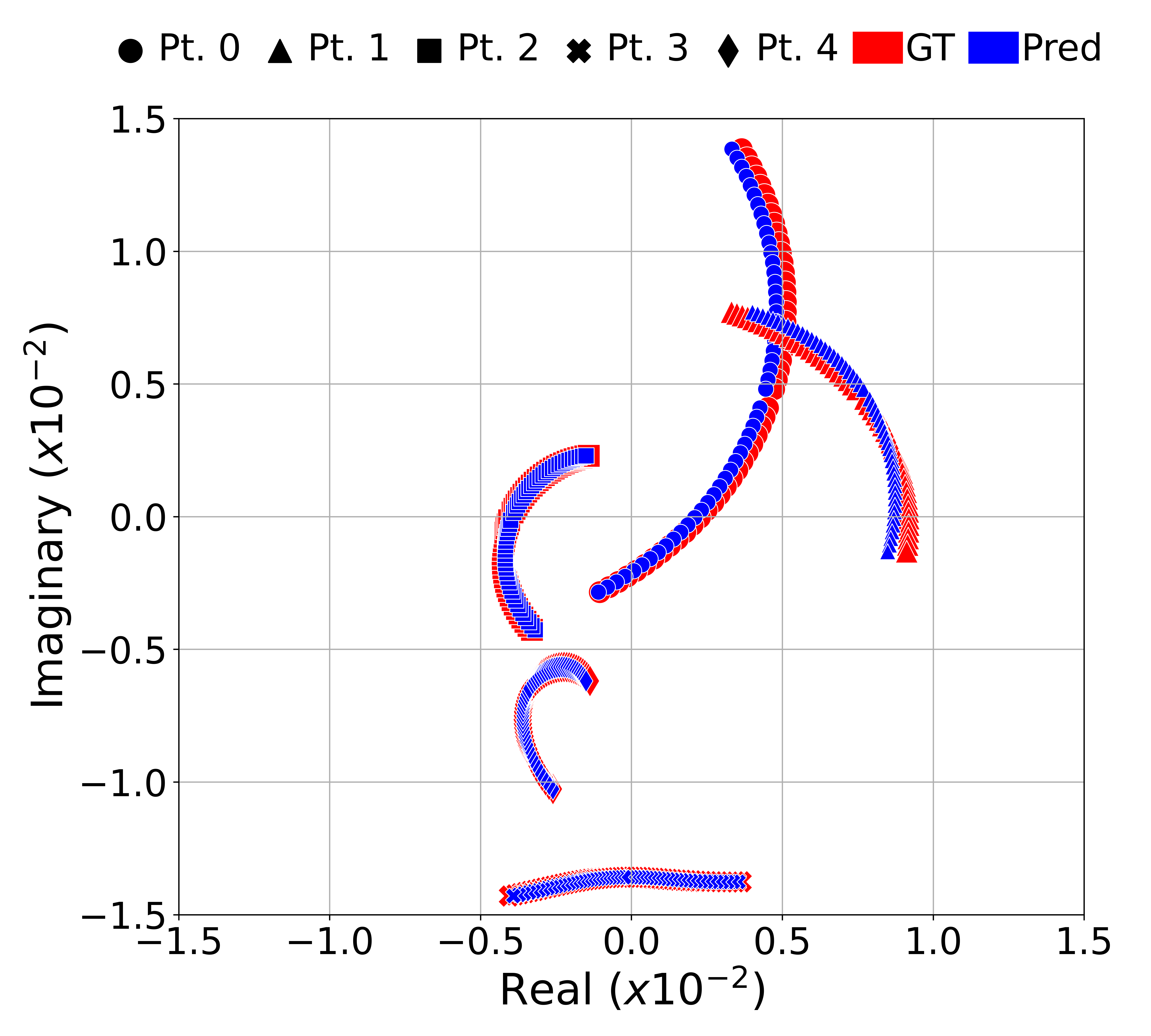}
        \caption{}
        \label{fig:subcarrier_constellation}
     \end{subfigure}
     \hfill
        \caption{\textbf{Subcarrier Channel Prediction}}
    \end{minipage}
\end{figure*}

\section{Subcarrier Channel Prediction}

% \begin{figure*}[ht]
%      \centering
%      \hfill
%      \begin{subfigure}{0.48\linewidth}
%         \centering
%         \includegraphics[width=\linewidth]{Figures/subcarriers.png}
%          \caption{}
%          \label{fig:subcarrier}
%      \end{subfigure}
%      \hfill
%     \begin{subfigure}{0.48\linewidth}
%          \centering
%         \includegraphics[width=\linewidth]{Figures/subcarriers_constellation.png}
%         \caption{}
%         \label{fig:subcarrier_constellation}
%      \end{subfigure}
%      \hfill
%         \caption{\textbf{Subcarrier Channel Prediction}}
% \end{figure*}

Modern wireless communication systems, such as WiFi and 5G use Orthogonal Frequency Division Multiplexing (OFDM) for modulation. OFDM divides the spectrum bandwidth into multiple narrow frequency bands, termed subcarriers. Each subcarrier modulates one data symbol. To demodulate data, the receiver has to perform channel estimation for all subcarriers. For instance, a 20~MHz WiFi channel is typically divided into 64 subcarriers, with each subcarrier 312.5~kHz apart from its neighbors. The channels of 52 (out of the 64) subcarriers are typically measured and reported by commercial off-the-shelf WiFi chips~\cite{ma2019wifi}, known as channel state information (CSI). 

\name is able to predict the subcarrier channels accurately using a model trained only on a single subcarrier without further fine-tuning. To demonstrate this capability, we use a model trained on the conference room dataset with carrier frequency 2.412~GHz to infer the channels for its nearby subcarriers. To synthesize the channel at a specific subcarrier frequency, we simply change the signal frequency $f$ in the channel synthesis algorithm (Eq.~\eqref{eq:rendering}) to the subcarrier frequency $f_{sc} = f_c+k\times \Delta f~Hz$, where $k$ is the index of the subcarrier, ranging from -26 to 26; $\Delta f$ is the subcarrier spacing, which equals to 3.125~kHz for 802.11 a/g/n; and $f_c$ is the carrier frequency of the channel. Figure~\ref{fig:subcarrier} shows the predicted amplitude and phase versus the subcarrier indices at five different receiver locations (Pt.0$\sim$ 4). As is shown, the predictions align very well with the ground truth. Figure~\ref{fig:subcarrier_constellation} shows a different view of the same data as~\ref{fig:subcarrier}, where we plot the real and imaginary parts of the subcarriers in x and y-axis, respectively. In particular, we note that the prediction accuracy remains high even for highly frequency selective channels. For instance, the channel amplitude of Pt.0 varies by orders of magnitude across different subcarriers. And \name is still able to predict the CSI accurately.

The reason behind this capability of \name is that the model learns to decouple the attenuation factor and phase rotation due to the reflection from that of free space propagation. The model predicts only the radiated signal, which is nearly unchanged across the subcarrier frequencies. The dominant cause of the varying patterns across subcarriers is the phase shift during the free space propagation and constructive/destructive interference of multipath, which are explicitly modeled by the channel synthesis algorithm.

\section{Impact of Noise}

Here, we are interested in seeing the robustness of \name against the influence of noise in wireless measurements. In this experiment, we add artificial Gaussian white noise to the simulated channels. The ratio between the channel power and added noise power is denoted as channel SNR. Figure~\ref{fig:ch_noise} shows the results, the blue line shows the results of adding a fixed noise, i.e. bias, to the ground truth channel in the training set. As is expected, the prediction accuracy reduces as the channel SNR decreases. However, typical wireless systems today achieve more than 20 dB channel SNR in practice for reliable communication, where \name performs well in such SNRs. Also, we note that collecting multiple measurements with different biases can improve the robustness. The orange line in Figure~\ref{fig:ch_noise} demonstrates this effect, where we use measurements with different additive noises in each iteration of training. The model is able to maintain good performance even at very low SNRs. This is because the model implicitly averages these measurements to reduce the bias during the training. 

%%%%%%%%%%%%%%%%%%%%%%%%%%%%%%%%%%%%%%%%%%%%%%%%%%%%%%%%%%%%%%%%%%%%%%%%%%%%%%%
%%%%%%%%%%%%%%%%%%%%%%%%%%%%%%%%%%%%%%%%%%%%%%%%%%%%%%%%%%%%%%%%%%%%%%%%%%%%%%%

\end{document}